\documentclass[iop,tighten]{emulateapj}

\usepackage{subfigure}

\newcommand{\be}{\begin{equation}}
\newcommand{\ee}{\end{equation}}

\newcommand{\msun}{{$M_{\odot}$}}

\newcommand{\gtsima}{$\; \buildrel > \over \sim \;$}
\newcommand{\ltsima}{$\; \buildrel < \over \sim \;$}
\newcommand{\prosima}{$\; \buildrel \propto \over \sim \;$}
\newcommand{\gsim}{\lower.5ex\hbox{\gtsima}}
\newcommand{\lsim}{\lower.5ex\hbox{\ltsima}}
\newcommand{\simgt}{\lower.5ex\hbox{\gtsima}}
\newcommand{\simlt}{\lower.5ex\hbox{\ltsima}}
\newcommand{\simpr}{\lower.5ex\hbox{\prosima}}
\newcommand{\es}{erg~s$^{-1}$}

\newcommand{\cxo}{\textit{Chandra}}
\newcommand{\hst}{\textit{Hubble}}
\newcommand{\spi}{\textit{Spitzer}}
\newcommand{\bhm}{$M_{\rm BH}$}
\newcommand{\lx}{$L_{\rm X}$}

\newcommand{\eddra}{$L/L_{\rm Edd}$}
\newcommand{\xeddra}{$L_{\rm X}/L_{\rm Edd}$}

\begin{document}

\title{AMUSE-Field I: Nuclear X-ray Properties of Local Field and
Group Spheroids across the Stellar Mass Scale}

\author{Brendan~Miller,$^{1}$ Elena~Gallo,$^{1}$ Tommaso~Treu,$^{2}$
and Jong-Hak~Woo$^{3}$}

\footnotetext[1]{Department of Astronomy, University of Michigan, Ann
Arbor, MI 48109, USA}

\footnotetext[2]{Physics Department, University of California, Santa
Barbara, CA 93106, USA}

\footnotetext[3]{Astronomy Program, Department of Physics and
Astronomy, Seoul National University, Seoul, Republic of Korea}

\begin{abstract}

We present the first results from AMUSE-Field, a \cxo~survey designed
to characterize the occurrence and intensity of low-level accretion
onto supermassive black holes (SMBHs) at the center of local
early-type field galaxies. This is accomplished by means of a
Large~Program targeting a distance-limited ($<$30~Mpc) sample of 103
early types spanning a wide range in stellar masses. We acquired new
ACIS-S observations for 61 objects down to a limiting (0.3--10~keV)
luminosity of $2.5\times 10^{38}$ \es, and we include an additional 42
objects with archival (typically deeper) coverage. A nuclear
\hbox{X-ray} source is detected in 52 out of the 103 galaxies. After
accounting for potential contamination from low-mass \hbox{X-ray}
binaries, we estimate that the fraction of accreting SMBHs within the
sample is $45\pm7$\%, which sets a firm lower limit on the occupation
fraction within the field. The measured nuclear X-ray luminosities are
invariably highly sub-Eddington, with $L_{\rm X}/L_{\rm Edd}$ ratios
between $\sim$$10^{-4}$--$10^{-8}$. As also found in a companion
survey targeting Virgo early types, the active fraction increases with
increasing host galaxy stellar mass, reflective of ``Eddington
incompleteness'' within the lower-mass objects. For the Field sample,
the average nuclear X-ray luminosity scales with the host stellar mass
as $M_{\rm star}^{0.71\pm0.10}$, with an intrinsic scatter of
$0.73\pm0.09$ dex. Qualitatively similar results hold for
morphologically homogeneous (type E) or uniform sensitivity (new
observations only) subsets. A majority of the AMUSE-Field galaxies
(78\%) inhabits groups, enabling us to investigate the influence of
group richness upon nuclear activity. We see no evidence for a
positive correlation between nuclear \hbox{X-ray} luminosity,
normalized to host properties, and galaxy density. Rather, while the
scatter is substantial, it appears that the Eddington-scaled
\hbox{X-ray} luminosity of group members may be slightly lower than
for isolated galaxies, and that this trend continues to cluster
early-types.

\end{abstract}


\section{Introduction}

Convincing evidence that supermassive black holes (SMBHs) are able to
form in the early universe is provided by observations of
high-redshift quasars (e.g., Volonteri \& Rees 2006; Vestergaard \&
Osmer 2009; Willott et al.~2010; Treister et al.~2011). The seeds for
these early SMBHs may have been produced from supermassive (e.g.,
Begelman 2010) or Population III stars, or from direct gas collapse
(e.g., Volonteri \& Natarajan 2009), perhaps in massive protogalaxy
mergers (e.g., Mayer et al.~2010; Volonteri 2010). While high-redshift
quasars generally display optical/UV and X-ray properties similar to
those of their local analogues (e.g., Shemmer et al.~2006), {\it
Spitzer\/} observations indicate many are young sources (some lack hot
dust) that are growing rapidly (Jiang et al.~2010). In contrast,
similarly massive SMBHs at more moderate redshifts are growing more
slowly (e.g., Netzer et al.~2007). Indeed, the growth of SMBHs appears
to be ``anti-hierarchical'' in the sense that active accretion is
concentrated in higher/lower-mass SMBHs at earlier/later cosmological
times (e.g., Heckman et al.~2004; Merloni \& Heinz 2007; Shankar et
al.~2009; Gallo et al.~2010; Goulding et al.~2010; Kelly et al.~2010;
Lamastra et al.~2010; Schulze \& Wisotzki 2010; Schawinski et
al.~2010).

The peak of the quasar space density around $z\sim2$ (e.g., Brown et
al.~2006; Hopkins et al.~2007; Kelly et al.~2010)\footnote{The quasar
peak redshift is luminosity-dependent, displaying a similar
``downsizing'' effect (e.g., Hasinger et al.~2005; Croom et al.~2009)
to that found for SMBH growth.} and the $\sim$10$^{8}$~yr quasar
lifetime (e.g., Yu \& Tremaine 2002) suggest (e.g., Soltan 1982) that
``inactive'' galactic nuclei typically also host SMBHs accreting at
low levels and/or radiating inefficiently in a post-quasar stage which
may result after much of the available fuel has been consumed or
expelled (e.g., Hopkins et al.~2005). The distinctions between active
galactic nuclei (AGNs), low-luminosity AGNs (Ho 1999, 2008), and
formally inactive galaxies are somewhat arbitrary, but as a general
guideline (which we will adhere to in this work) AGNs display
bolometric, Eddington-scaled luminosities in excess of a few percent,
low-luminosity AGNs are in the range $10^{-4}<L/L_{\rm Edd}<10^{-2}$,
and inactive nuclei are highly sub-Eddington, with \eddra \simlt
$10^{-4}$.  The Milky Way is one example of a formally inactive
galaxy. It is known to host a quiescent central SMBH with mass of
$3.6\times10^{6} M_{\odot}$ (e.g., Sch{\"o}del et al.~2009) from which
low-level, persistent accretion-powered \hbox{X-ray} emission has been
detected (\xeddra$\simeq 10^{-11}$; Baganoff et al.~2001). Even low
levels of SMBH accretion-powered activity in nearby galaxies may be
detected efficiently in sensitive, high spatial resolution
\hbox{X-ray} observations, although contamination from bright
\hbox{X-ray} binaries must be properly accounted for when dealing with
nuclear X-ray luminosities comparable to the Eddington limit for a few
solar masses (e.g., Zhang et al.~2009; Gallo et al.~2010). This
problem can be substantially alleviated by restricting the search for
highly sub-Eddington nuclei to early-type galaxies, as they
conveniently avoid the brighter high-mass \hbox{X-ray} binaries
produced concurrently with star formation (e.g., King et al.~2001;
Ghosh et al.~2009).

From \hbox{X-ray} and other observations as well as numerical
simulations, a coherent scenario for the history and observed
characteristics of quiescent or weakly accreting SMBHs in early-type
galaxies is emerging. Although hot halo gas or stellar winds provide
readily available fuel and gas infall may proceed at near the Bondi
rate, outflows can drive off a substantial fraction of the accreting
mass and the radiative efficiency is inferred from observed spectral
energy distributions to be highly sub-Eddington (e.g., Di Matteo et
al.~2000; Pellegrini 2005; Soria et al.~2006b).

The growth of SMBHs is believed to be closely tied to the evolution of
their hosts (as evidenced by, for example, the correlations between
SMBH mass and bulge luminosity or central stellar velocity dispersion;
see, e.g., overview by Ferrarese \& Ford 2005). Radiative and
mechanical AGN feedback act to regulate SMBH accretion and quench star
formation (e.g., Ciotti et al.~2009, 2010; Pellegrini et al.~2012),
and mechanical feedback continues to play an important role in
low-activity ``radio-mode'' systems (e.g., Croton et al.~2006; Merloni
\& Heinz~2007; Marulli et al.~2008). The assembly of early-type galaxies 
apparently proceeds hierarchically, despite the older ages and shorter
star-formation timescales of massive ellipticals (e.g., De Lucia et
al.~2006; Eliche-Moral et al.~2010), with cold-gas-rich/poor
(``wet''/``dry'') mergers tending to produce discy/boxy types (e.g.,
Kang et al.~2007; Hopkins \& Quataert~2011).

Naturally the properties of higher-mass SMBHs (in more massive
galaxies) may be more easily constrained; the dominance of compact
stellar nuclei at lower luminosities (e.g., Ferrarese et al.~2006a)
presented initial observational challenges to establishing whether
lower-mass early-type galaxies necessarily even contained a SMBH. The
presence of ``missing light'' or cores within luminous ellipticals may
be explained by binary black holes disrupting the interior region in
the process of merging (e.g., Kormendy et al.~2009 and references
therein), and so preserved nuclear star clusters could potentially
indicate a lack of a SMBH binary phase (e.g., Kang et
al.~2007). However, it is now clear that SMBHs may coexist with
nuclear star clusters (e.g., Seth et al.~2008; Graham \& Spitler~2009;
Kormendy et al.~2009).

To quantify the rate of low-level SMBH activity over a well defined
sample of nearby early-type galaxies, our group carried out the AGN
MUltiwavelength Survey of Early-Type Galaxies in the Virgo
cluster. AMUSE-Virgo (ID 8900784, \cxo~Cycle~8, PI: Treu, 454~ks)
targeted the 100 spheroidal galaxies which compose the \hst~ACS Virgo
Cluster Survey (VCS; C{\^o}t{\'e} et al.~2004) with
\cxo~ACIS-S and \spi~MIPS (Gallo et al.~2008, 2010, hereafter G08, G10; 
Leipski et al.~2012). The VCS sample was selected based
solely on optical properties and spans a wide range in stellar mass,
from $10^{8}-10^{12}~M_{\odot}$. At the average Virgo distance of
16.5~Mpc (Mei et al.~2007), the AMUSE-Virgo {\it Chandra\/} snapshot
observations (about 5 ks each) reached a $2\sigma$ limiting 0.5--7~keV
luminosity of $1.3\times10^{38}$~erg~s$^{-1}$ (i.e., close to the
Eddington limit for 1 \msun). A nuclear \hbox{X-ray} source was
detected by G10 in 32/100 objects, which, after taking into
consideration the minor but non-negligible effects of low-mass
\hbox{X-ray} binary (LMXB) contamination, sets a lower limit of
24--34\% (at the 95\% confidence level) to the occupation fraction of
SMBHs in the nuclei of cluster early-type galaxies. Within the
AMUSE-Virgo survey the fraction of
\hbox{X-ray}-active SMBHs is seen to increase with host stellar
mass,\footnote{That SMBH activity in general is detected more
frequently with larger host stellar mass or luminosity is known since
the first Palomar sample (e.g., Ho et al.~1997).} but in a manner
consistent with arising from {``Eddington incompleteness"}, i.e., the
intrinsic inability of any luminosity-limited survey to reach the same
Eddington-scaled luminosity level across a wide range of inferred
black hole masses. Despite this effect, which obviously penalizes
lower mass black holes, the average Eddington-scaled \hbox{X-ray}
luminosity of the AMUSE-Virgo galaxies is found to scale inversely
with black hole mass as $\langle L_{\rm X}/L_{\rm Edd}\rangle \propto
M_{\rm BH}^{-0.62}$, so that nearby lower-mass SMBHs are relatively
more \hbox{X-ray}-active {(a trend G10 refer to as ``downsizing in
black hole accretion'')}. In all cases the \hbox{X-ray} luminosity is
highly sub-Eddington, with $10^{-8}{\simlt}L_{\rm X}/L_{\rm
Edd}{\simlt}10^{-5}$.

While this and other studies (e.g., Pellegrini 2005; Zhang et
al.~2009; Boroson et al.~2011) have provided useful and detailed
census of low-level SMBH activity within the local universe, including
as a function of the host galaxy properties (Pellegrini 2010), {\it
the impact of environment upon the occurrence and intensity of SMBH
activity remains unclear}. The properties of early-type galaxies in
more sparsely populated regions are distinct from their counterparts
in clusters: relative to cluster sources, field early-type galaxies
face reduced ram pressure stripping (e.g., Acreman et al.~2003;
Gavazzi et al.~2010; Shin et al.~2012) and on average contain more
cold gas and tend to have younger stellar populations (e.g., Thomas et
al.~2005; Oosterloo et al.~2010). Within clusters, tidal interactions
play an important role (e.g., Gnedin 2003), for example as harassment
from high-speed encounters (e.g., Treu et al.~2003); such effects are
only infrequently relevant to isolated field galaxies, although within
groups the lower galaxy velocity dispersions can help facilitate
mergers. Early work with {\it ROSAT\/} links the environment with
\hbox{X-ray} emission, but to a debated degree; for example, Brown \&
Bregman (2000) found that cluster early-type galaxies tend to be more
\hbox{X-ray} luminous than their field counterparts, and hypothesized
that this results from environmental suppression of outflows or the
availability for accretion of hot intergalactic gas within clusters,
whereas O'Sullivan et al.~(2001) found similar $L_{\rm X}(L_{\rm B})$
relations for cluster, group, and field galaxies. While such results
are intriguing, their utility is necessarily restricted by the limited
angular resolution of {\it ROSAT\/}, which makes it difficult to
disentangle diffuse thermal
\hbox{X-ray} or off-nuclear \hbox{X-ray} binary emission from that
linked to the SMBH.

The goal of this work is to investigate low-level SMBH activity within
a well defined sample of nearby field spheroidal galaxies (E and E-S0
galaxies), and to characterize the dependence (if any) of such
activity upon environmental effects, such as living in isolation
versus belonging to a group of galaxies, or belonging to a poor versus
a rich group. Toward this end, we were awarded a Large {\it Chandra\/}
program building upon and extending AMUSE-Virgo to non-cluster
environments: AMUSE-Field (ID 11620915, \cxo~Cycle 11, 479~ks, PI
Gallo) carried out new observations of 61 field early-type galaxies
located within 30~Mpc and spanning a wide range of stellar and black
hole masses. The new observations are supplemented with an additional
42 objects with archival coverage, for a total of 103 objects.

This paper is structured as follows: $\S$2 describes the sample
selection and {\it Chandra\/} data reduction; $\S$3 provides the
\hbox{X-ray} results; $\S$4 investigates the dependence of nuclear
\hbox{X-ray} luminosity on stellar mass; $\S$5 discusses environmental
influences upon SMBH accretion; $\S$6 summarizes and
concludes. $L_{\rm X}$, $M_{\rm star}$, and $M_{\rm BH}$ are the
nuclear 0.3--10~keV \hbox{X-ray} luminosity, galaxy stellar mass, and
black hole mass, respectively, and are given in units of erg~s$^{-1}$,
$M_{\odot}$, and $M_{\odot}$ throughout. Errors are $1\sigma$ unless
otherwise noted.

\section{Sample selection and {\it Chandra\/} data reduction}

The AMUSE-Field sample was selected based only on optical galaxy
properties, in the following manner. We first searched the
HyperLeda\footnote{http://leda.univ-lyon1.fr/} catalog (Paturel et
al.~2003) for spheroidal galaxies, conservatively defined as having
type E or E-S0, with inferred heliocentric velocities less than
1800~km~s$^{-1}$ and distances less than 30~Mpc.\footnote{For
significantly larger distances, the total stellar mass enclosed by the
{\it Chandra\/} point spread function becomes sufficiently large that
the possibility of cumulative LMXB contamination is non-negligible;
see discussion in G10.} Objects near the Galactic plane
($|b|<30^{\circ}$) were excluded, to avoid absorption effects. Objects
near the Virgo cluster (within $\simeq6^{\circ}$ of 12.44h,
$12.72^{\circ}$) or the Fornax cluster (within $\simeq0.2^{\circ}$ of
3.64h, $-35.45^{\circ}$) were likewise excluded, to avoid
contamination by cluster galaxies; these are the only clusters lying
within the considered volume. Finally, objects were required to have
absolute $B$ magnitudes more luminous than $M_{\rm B}=-13$. This
results in a list of 204 objects. Of these 204 objects, 42 already
possessed high-quality archival {\it Chandra\/} ACIS-S3 imaging
observations, which we define here as exposure $>3$~ks and off-axis
angle $<5'$. 61 additional targets were selected such that the full
AMUSE-Field sample, made up of 103 early-type galaxies, provides
representative coverage of a wide range in absolute $B$
magnitude. \cxo~exposures for the 61 new targets (which we shall
generally refer to as ``snapshots'') vary between $\sim 2-12$ ks, and
were chosen to achieve a common sensitivity threshold of $2.5\times
10^{38}$~\es~\hbox{(0.3--10 keV)}. Optical properties of the
individual galaxies\footnote{The full AMUSE-Field sample includes four
objects for which we had \hbox{X-ray} data in hand but that do not
technically meet all selection criteria. NGC 5077 has a distance of
40.2~Mpc but is otherwise suitable. PGC 132768, 042737, and 042596
(all \hbox{X-ray} non-detections) do not have a morphological type
listed in HyperLeda; digital sky images suggest their morphologies,
while spheroidal, are somewhat irregular.} are listed in Table~1,
along with their derived stellar and black hole masses ($\S$4.1) and
their group membership ($\S$5).

The {\it Chandra\/} ACIS-S \hbox{X-ray} data were reduced in a uniform
fashion following the general procedure applied to the AMUSE-Virgo
sample (G08, G10). Here we describe the astrometry correction, event
reprocessing, lightcurve filtering, source detection, and aperture
photometry. The calculation of \hbox{X-ray} luminosities from net
count rates is described in $\S$3.1.

The default {\it Chandra\/} astrometry is generally accurate to
$\simlt0.3''$ (1$\sigma$; e.g., $\S$3 of Eckart et al.~2005), which is
sufficient to identify uniquely which \hbox{X-ray} source, if any, is
located nearest the galactic nucleus. For cases in which the pointing
fell within the SDSS footprint, we improve the \hbox{X-ray} astrometry
through cross-matching \hbox{X-ray} sources to their optical
counterparts. A first-pass \hbox{X-ray} source list was created by
running {\it wavdetect\/} on a \hbox{0.3--7}~keV image generated from
the pipeline level~2 event file. For each non-nuclear ($>10''$)
ACIS-S3 \hbox{X-ray} source, the nearest primary SDSS object with
$r$-band magnitude $m_{\rm r}<23$ (hence highly accurate optical
position) was sought, to within a radius dependent upon the {\it
wavdetect\/} estimated net counts as well as the \hbox{X-ray} off-axis
angle (this \hbox{X-ray} positional uncertainty is based on $\S$2.3 of
Hong et al.~2005). The average offset in right ascension and
declination was calculated using weighting by the inverse
uncertainties; any 3$\sigma$ outliers were discarded and this process
was repeated. The resulting offsets were converted to pixel shifts and
used to generate a new aspect solution file.  The number of ``clean''
matches to SDSS objects used to calculate the shifts ranged from 2 to
10, and in no cases were the shifts in $x$ or $y$ greater than one
pixel (i.e., 0.5$''$). Astrometry for pointings outside the SDSS
footprint, or for observations lacking sufficient high-confidence
cross matches, was not adjusted.

A background lightcurve of the S3 chip (with point sources excluded),
binned to 200~s, was screened for flaring using the {\it deflare\/}
script. Periods of anomalously high or low background were identified
as any $>3\sigma$ deviations from the mean count rate for most
observations, and as any $>2.5\sigma$ or $>2\sigma$ deviations in a
few instances where strong flaring was present, and confirmed through
manual inspection. Such periods were then excluded through application
of the {\it deflare\/}-generated good-time intervals as a filter to
the level 2 event file.

An updated point source list was then compiled through running {\it
wavdetect\/} on a clean 0.3--7~keV image with scales of 1, 1.4, 2,
2.8, and 4 pixels, using a 1.5~keV exposure map and a threshold
significance of 10$^{-6}$ (which corresponds to approximately one
false detection expected per chip). Net count rates were calculated
for each source from aperture photometry conducted within 95\%
encircled energy radii (at 1.5~keV), with local background estimated
from the median of eight nearby regions. For Field galaxies lacking a
nuclear \hbox{X-ray} detection, the 95\% confidence upper limit is
estimated from the local background. Where the number of background
counts is less than 10 we use the Bayesian formalism of Kraft et
al.~(1991) to determine the limit; elsewhere, we use Equation 9 from
Gehrels (1986).

As they are the brightest galaxies within our volume-limited sample,
the archival sources are generally more luminous, more massive, and
thus able to retain more hot gas.  For those objects, it is
advantageous to determine the nuclear SMBH \hbox{X-ray} luminosity
from the hard-band counts, so as to minimize contamination from the
diffuse soft emission. The diffuse emission may be modeled as thermal
(hot gas) and power-law (unresolved LMXBs) components (e.g., G10;
Boroson et al.~2011). For these galaxies, the contribution from hot
gas generally drops below $\sim$5\% of the nuclear \hbox{X-ray} flux
at energies $\simlt$2~keV. Therefore, for archival galaxies with
sufficiently high signal-to-noise data, defined as $\ge50$ counts in
the 0.3--7~keV band (29/42 objects), we calculate \hbox{X-ray} flux
from the 2--7~keV count rate.\footnote{There are two exceptions: one
snapshot object, ESO 540-014, has sufficient counts to permit use of
the 2--7~keV rate, and one archival object, NGC~1052, has a
sufficiently hard observed spectrum that we use the 0.3--7~keV rate to
avoid overestimating the \hbox{X-ray} flux.} In a few instances for
which strong diffuse emission overlapped the nucleus even in the hard
band, the local background was manually adjusted.

\begin{figure}
\includegraphics[scale=0.45]{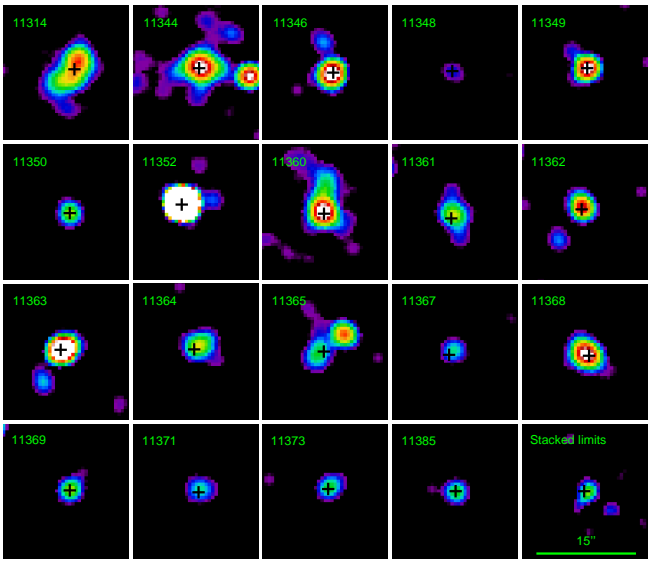}
\figcaption{\small ACIS-S 0.3--7 keV images of those snapshot targets
(labeled by ObsID; see Table~2) with detected nuclear X-ray emission,
smoothed with a Gaussian kernel of five pixels ($2.5''$) and plotted
with common squareroot intensity scaling. Images are centered at the
catalog optical coordinates of the host galaxy, and the position of
the X-ray source associated with the nucleus is marked with a black
cross. The rightmost panel in the bottom row shows the stacked image
of snapshot targets that individually lack a nuclear X-ray detection
(scaling does not match other frames); the central region is detected,
with an average X-ray luminosity consistent with arising from a
population of enclosed LMXBs.}
\end{figure}

\section{Results}

\subsection{Nuclear X-ray census}

X-ray properties of the AMUSE-Field sample are listed in Table~2, and
notes on selected archival objects are given in Appendix~A. From the
103 field galaxies, a nuclear \hbox{X-ray} source is detected in 52
cases, for a detection fraction of 50\%. The detection fraction is
naturally higher for the archival objects (33/42, or 79\%), due to
their generally deeper exposures\footnote{Of the nine archival upper
limits, five have effective exposure times less than 10~ks, two are
small galaxies serendipitously observed off-axis ($\theta\simgt2.5'$),
and two (NGC 5846 and NGC 3923; see Appendix~A) do not show a cleanly
resolved central point source.} References for previously published
analysis of the archival objects are also listed in Table~2.

The detection fraction for the snapshot objects is 19/61, or
31\%. Cutout images of all snapshot targets with detected nuclear
\hbox{X-ray} emission are shown in Figure~1, smoothed with a Gaussian
kernel of five pixels.  Count rates (either 0.3--7~keV or 2--7~keV, as
discussed in $\S$2) were converted to 0.3--10~keV fluxes with
PIMMS,\footnote{http://cxc.harvard.edu/toolkit/pimms.jsp} assuming a
power-law spectrum with $\Gamma=2$ and Galactic absorption. For each
galaxy the radial-velocity distance modulus in HyperLeda was used to
calculate the nuclear 0.3--10~keV \hbox{X-ray} luminosity, $L_{\rm
X}$. For most of these objects, non-redshift distances are not
available in HyperLeda or NED; from the remainder, the non-redshift
distance modulus from HyperLeda is an average of 0.16 lower,
corresponding to a distance $\sim1.5$~Mpc closer at $d\sim20$~Mpc.
For the low redshift of these objects, the $k$-correction is
insignificant and is therefore neglected.

Nine field galaxies have $\log $\lx \simgt$39.8$: NGC 1052, 2768,
4036, 4203, 4278, 4494, 5077, IC 1459, and ESO 540$-$014. All but the
last of these are from the archival sample. Each one of these nine
galaxies has radio and/or optical properties consistent with the
presence of low-level AGN activity. NGC 1052, 4036, 4278, 4494, and IC
1459 are included as LINER/AGNs in Gonz{\'a}lez-Mart{\'{\i}}n et
al.~(2009); NGC 1052, 4278, 5077, and IC 1459 are categorized as
flat-spectrum radio sources in Healey et al.~(2007); NGC 2768, 4036,
4203, and ESO 540$-$014 are identified as active in V{\'e}ron-Cetty \&
V{\'e}ron~(2006) as Seyfert~1, LINER~b, LINER~b, and Seyfert~2
galaxies, respectively. In comparison, the AMUSE-Virgo sample contains
only two objects (M87 and NGC 4564) with $\log{L_{\rm X}}>39.8$ (from
Table~1 of G10).

\subsection{Origins of the nuclear X-ray emission}

The correspondence between higher values of \hbox{X-ray} luminosity
and optical or radio indicators of activity ($\S$3.1) already supports
that, for at least these objects, the nuclear \hbox{X-ray} emission is
linked to the central SMBH. Indeed, prior studies of early-type
galaxies (G08, G10, and references in $\S$1) have overwhelmingly
associated nuclear point-source $L_{\rm X}$ measurements with
low-level SMBH activity. While our observations are not designed to
determine the precise \hbox{X-ray} emission mechanism, various
proposed inefficient accretion flow models (see, e.g., discussion and
references in Soria et al.~2006b) may apply to the objects in the
AMUSE samples. We here also consider alternative origins for the
nuclear \hbox{X-ray} emission, and find that they likely explain at
most a small fraction of the observed detections.

The probability of finding an unrelated background source of equal or
greater flux within the {\it Chandra\/} PSF for any given nuclear
\hbox{X-ray} detection is negligible, based on the $N(S)$ relation
provided by Moretti et al.~(2003), which Bauer et al.~(2004) find
matches the source count distribution within the {\it Chandra\/} Deep
Fields. Contamination from LMXBs, while still unlikely, must be
considered as a potential source of nuclear \hbox{X-ray} emission
($\S$3.2.1). Tidal disruption of stars may provide an alternative
method of fueling SMBHs ($\S$3.2.2).

\subsubsection{Low mass X-ray binary contamination}

Where a nuclear star cluster is not present, the total number of LMXBs
and their cumulative \hbox{X-ray} luminosity scale approximately with
the stellar mass (Gilfanov 2004; Kim \& Fabbiano 2004; Humphrey \&
Buote 2006), and so the number of LMXBs per unit stellar mass above a
particular luminosity threshold may be estimated from the \hbox{X-ray}
luminosity function for LMXBs (e.g., Gilfanov 2004). We estimate the
projected stellar mass enclosed by the {\it Chandra\/} PSF based on
the fractional luminosity within 2$''$; as most of the Field galaxies
currently lack high-resolution {\it HST\/} imaging, we use as a
comparative template similarly-distant Virgo early-type galaxies with
$-17<M_{\rm B}<-16$ (i.e., $\log{(M_{\rm star}/M_{\odot})}\sim9-10$)
which tend to have effective radii 10--20$''$ with Sersic indices of
1.0--2.5 (Ferrarese et al.~2006b). The total \hbox{X-ray} luminosity
from LMXBs within the nucleus of a typical such galaxy is $L_{\rm
X}=1.0\times10^{37}$~erg~s$^{-1}$, which is 4\% of the Field detection
limit. The expected nuclear contribution from LMXBs can be several
times greater in large ellipticals, where the total stellar mass can
be $\log{(M_{\rm star}/M_{\odot})}\sim11$ (here the smaller fraction
of the effective radius enclosed by the PSF is partially offset by a
higher Sersic index, for simple models, but see, e.g., Ferrarese et
al.~2006b for discussion of surface brightness profiles), but for such
galaxies in the Field sample the nuclear \hbox{X-ray} luminosities are
generally substantially above the detection limit. In summary, due to
the steepness of the LMXB luminosity function above a few
10$^{38}$~erg~s$^{-1}$, for most of the Field galaxies it is highly
unlikely that the PSF-enclosed stellar mass would generate significant
LMXB contribution to the measured $L_{\rm X}$ values. Figure~2 shows
the $\log{L_{\rm X}}$ number density for the Field sample, which may
be satisfactorily modeled as $dN/dlogL_{\rm X}=100\times{(L_{\rm
X}/10^{38})^{-0.6}}$. As for the Virgo sample (see Figure~6 of G10),
this is much shallower than would be expected if most of the
\hbox{X-ray} detections were actually LMXB-dominated.

\begin{figure}
\includegraphics[scale=0.46]{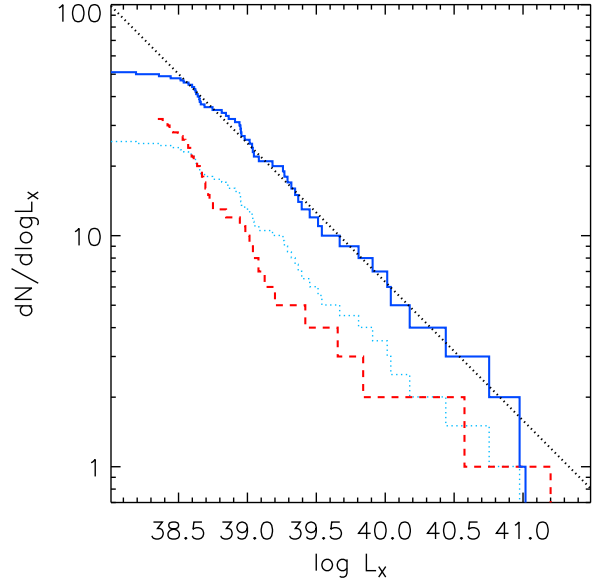} 

\figcaption{\small The $\log{L_{\rm X}}$ number density for the
AMUSE-Field sample (solid blue line) may be modeled as $dN/dlogL_{\rm
X}=100\times{(L_{\rm X}/10^{38})^{-0.6}}$ (dotted black line). For
comparison, the result for the AMUSE-Virgo sample (dashed red line) is
plotted along with the Field number density rescaled by 0.5 (dotted
cyan line). The luminosity function for LMXBs is much steeper (see,
e.g., Figure 6 of G10), confirming that in almost all cases the
\hbox{X-ray} detections are linked to low-level SMBH activity.}
\end{figure}

Massive nuclear star clusters, with inferred radii around a few tens
of pc, become increasingly prominent down the mass function (e.g.,
Bekki \& Graham 2010; Graham et al.~2011; and references therein). The
enhanced stellar encounter rates within a nuclear star cluster implies
a larger possibility of chance LMXB contamination. To account for this
effect, for nucleated galaxies G10 conservatively adopted the higher
\hbox{X-ray} luminosity function of LMXBs in globular clusters, as
estimated by Sivakoff et al.~(2007). Unfortunately, most of the field
galaxies discussed here currently lack the two-color high-resolution
optical imaging (i.e., carried out with {\it HST\/} ACS) necessary to
confirm the presence of a nuclear star cluster and to infer its mass.

To estimate the degree of LMXB contamination within the field sample
detections, we note that the AMUSE-Virgo sample contained 6/32 objects
with both a nuclear star cluster and an \hbox{X-ray} detection. All
six of these objects had $\log{(M_{\rm star}/M_{\odot})}<10.5$ (and in
total there are 16 Virgo \hbox{X-ray} detections with $\log{(M_{\rm
star}/M_{\odot})}<10.5$). The influence of nuclear star clusters at
higher stellar masses, even if present, is of decreasing relevance;
Graham \& Spitler (2009) find that in mixed nuclei the SMBH mass
dominates that of the nuclear star cluster for spheroid masses above
$\sim10^{10} M_{\odot}$. In 4/6 of the Virgo objects with both a
nuclear star cluster and an \hbox{X-ray} detection (i.e., in 4/16 of
Virgo \hbox{X-ray} detected objects with $\log{(M_{\rm
star}/M_{\odot})}<10.5$), G10 calculate that the \hbox{X-ray} emission
was likely due to LMXB rather than SMBH activity. Applying a similar
rate\footnote{We have been awarded Cycle 19 {\it HST\/}/ACS
observations to verify that the incidence of nuclear star clusters
within field early-types is similar to that found for the AMUSE-Virgo
survey.} of 4/16=0.25 to the 22 Field galaxies with $\log{(M_{\rm
star}/M_{\odot})}<10.5$ and \hbox{X-ray} detections suggests that in
$\simlt6$ cases the \hbox{X-ray} emission might be contaminated by a
bright LMXB; Figure~4 shows one plausible distribution of intrinsic
SMBH detections in the Field sample (here the galaxies with presumed
nuclear star cluster LMXB contamination all have $L_{\rm
X}<5\times10^{38}$~erg~s$^{-1}$). The detected intrinsic fraction of
accreting SMBHs within the Field sample is then $\sim$46/103 (rather
than the observed 52/103 rate of \hbox{X-ray}-detected nuclei), or
$45\pm7$\%, which sets a lower limit to the occupation fraction of
SMBHs within these galaxies.

Off-nuclear \hbox{X-ray} sources from many of the archival
observations are discussed elsewhere (e.g., Irwin et al.~2003). Within
the snapshot observations, several off-nuclear \hbox{X-ray} sources
are detected within three times the effective radius (this limit is
sometimes used to construct catalogs of ultraluminous \hbox{X-ray}
sources (ULXs) or LMXBs; e.g., Irwin et al.~2003). Based on the cosmic
$N(S)$ relation previously discussed, the majority of these are
unlikely to be unrelated background sources. Presuming the off-nuclear
\hbox{X-ray} sources have distances as for the corresponding galaxy,
we do not find any ULXs with $L_{\rm X}\simgt10^{40}$~erg~s$^{-1}$. A
complete catalog of ULXs for the Field snapshot galaxies will be
presented in a later work, in which their properties will be placed in
context with our upcoming {\it HST\/} observations and will be
compared to ULXs found within the AMUSE-Virgo snapshot sample.

\subsubsection{X-ray emission from stellar tidal disruption}

One interesting possibility for fueling SMBHs is through tidal
disruption of stars (e.g., Rees 1988). Simulations indicate that such
events could result in temporary super-Eddington infall, followed by
thin-disk accretion, likely transitioning to a radiatively-inefficient
flow (Strubbe \& Quataert 2009). The predicted spectrum includes soft
\hbox{X-ray} blackbody emission, plausibly with an additional
power-law tail (Strubbe \& Quataert 2011). A handful of individual
cases of potential tidal disruption events have been identified (e.g.,
Lin et al.~2011 and references therein), including most recently the
spectacular Swift J164449.3+573451 (Burrows et al.~2011; Levan et
al.~2011); in this rare case $L_{\rm X}$ significantly exceeds $L_{\rm
Edd}$ for the inferred modest black hole mass ($\log{M_{\rm
BH}/M_{\odot}}=5.5\pm1.1$; Miller \& G{\"u}ltekin~2011), suggesting
the observed emission comes predominantly from a beamed relativistic
jet (Bloom et al.~2011; Burrows et al.~2011; Cannizzo et
al.~2011). Deep surveys generally provide only loose upper limits on
the typical rate (e.g., $<10^{-4}$~yr$^{-1}$ per galaxy; Luo et
al.~2008), but numerical simulations carried out by Brockamp et
al.~(2011) suggest that approximately $3.0\pm1.4$ events per
10$^{5}$~yr are typical for a SMBH of 10$^{6} M_{\odot}$. We use this
rate to estimate the fraction of time for which a $10^6 M_{\odot}$
black hole would display $L_{\rm
X,tidal}>2.5{\times}10^{38}$~erg~s$^{-1}$ (the AMUSE-Field detection
limit). 

The timescale until the accretion rate drops below Eddington,
following the canonical $t^{-5/3}$ scaling (Rees 1988), is
$\simlt$0.3--1~yr (Strubbe \& Quataert 2009). Subsequent evolution of
the disk-dominated emission as $t^{-1.2}$ (Cannizzo et al.~1990) is
likely curtailed after $\dot{m}{\simlt}10^{-2}\dot{m_{\rm Edd}}$
(Strubbe \& Quataert 2009), around $\sim$100~yr, after which material
likely accretes through a thick disk. Presuming isotropic emission for
simplicity, and taking $L_{\rm X,tidal}{\simeq}10^{42}$~erg~s$^{-1}$
during the initial flare (motivated by Strubbe \& Quataert 2011), then
$L_{\rm X,tidal}{\simeq}4{\times}10^{39}$~erg~s$^{-1}$ after
$\sim$100~yr, dropping below our detectability threshold by
$\simlt$1000~yr (or sooner for an abrupt transition to radiatively
inefficient accretion). The fraction of time for which a tidal
disruption event could dominate the \hbox{X-ray} emission observed in
a similar AMUSE-Field galaxy is then $\simlt$3\%. (These estimates are
for a solar-mass main-sequence star; see, e.g., Li et al.~2002 and
Lodato et al.~2009 for alternatives). For reference, the AMUSE-Field
sample contains 31 galaxies with $M_{\rm BH}<10^{6} M_{\odot}$, so we
might anticipate detecting \hbox{X-ray} emission associated with
stellar disruption in about one object.

For SMBHs with $M_{\rm BH}>10^{7} M_{\odot}$, tidal disruption is
expected to be less significant, as the distance at which a stellar
trajectory can dynamically enter the loss cone exceeds the radius of
influence (Brockamp et al.~2011), and in addition the refill timescale
is quite long in large ellipticals, particularly core galaxies
(Merritt \& Wang 2005). Most (42/52) of the AMUSE-Field galaxies with
\hbox{X-ray} detections have $M_{\rm BH}>10^{7} M_{\odot}$. In view of
the above, we conclude that the nuclear \hbox{X-ray} emission in the
AMUSE-Field galaxies is not related to tidal disruption events,
although without multi-epoch observations we cannot rule it out for
any specific object. Instead, the SMBHs in these galaxies are more
likely currently fed through quasi-continuous but highly sub-Eddington
accretion, as arising from, for example, mass loss from evolved stars
(Soria et al.~2006b; Volonteri et al.~2011).

\subsection{Stacking non-detection}

The 42 snapshot observations lacking a nuclear \hbox{X-ray} detection
were stacked to check for a joint detection and assess their average
\hbox{X-ray} luminosity. After excising all off-nuclear \hbox{X-ray}
point sources (occupying areas here defined by their {\it wavdetect\/}
ellipses), the 0.3--7 keV images were stacked at the optical position
of each target. The combined image (Figure~1, lower right) has an
effective exposure of 307~ks. There is a significant excess of counts
near the center, with 42.2 net counts (87.1 total, 44.9 background)
present within a 10 pixel (4.9$''$) extraction radius. (Stacking
random positions does not produce a significant detection.) For an
average $N_{\rm H}$ of $2\times10^{20}$~cm$^{-2}$ and taking
$\Gamma=2$, the rate of $(1.4\pm0.4)\times10^{-4}$~counts~s$^{-1}$
corresponds to an average (unabsorbed) 0.3--10~keV flux of
1.1$\times10^{-15}$~erg~s~cm$^{-2}$. The mean distance modulus for
these 42 objects is 31.5 (20.0 Mpc), which gives an average
\hbox{X-ray} luminosity of $\log{L_{\rm X}}=37.7$~erg~s$^{-1}$. Given
the total enclosed stellar mass, this is broadly consistent with
arising from a population of individually undetectable LMXBs.

The excess \hbox{X-ray} emission within the stacked image is centrally
concentrated but not pointlike, and extends to $\simgt20''$. The net
rates within apertures of 4, 6, 10, 20, 40, and 60 pixels are 3.6,
6.3, 13.7, 23.0, 35.3, and $37.6\times10^{-5}$~counts~s$^{-1}$,
respectively. This indicates that the excess \hbox{X-ray} emission is
not exclusively nuclear, and likely includes contributions from
off-nuclear LMXBs as well as hot gas.

These results further confirm that the nuclear \hbox{X-ray}
luminosities calculated for snapshot targets with \hbox{X-ray}
detections do not suffer significantly from LMXB or hot gas
contamination. With the caveat that the presence or absence of a
nuclear star cluster has not been established for specific sources,
the average overall contribution to $L_{\rm X}$ from an underlying
ensemble of LMXBs within a 2--3$''$ aperture based on the above
measurement is $<$5--9\% of the minimum snapshot detected nuclear
\hbox{X-ray} emission ($\log{L_{\rm X}}=38.4$~erg~s$^{-1}$), and the
contribution from hot gas is negligible.

\begin{figure*}
\includegraphics[scale=0.9]{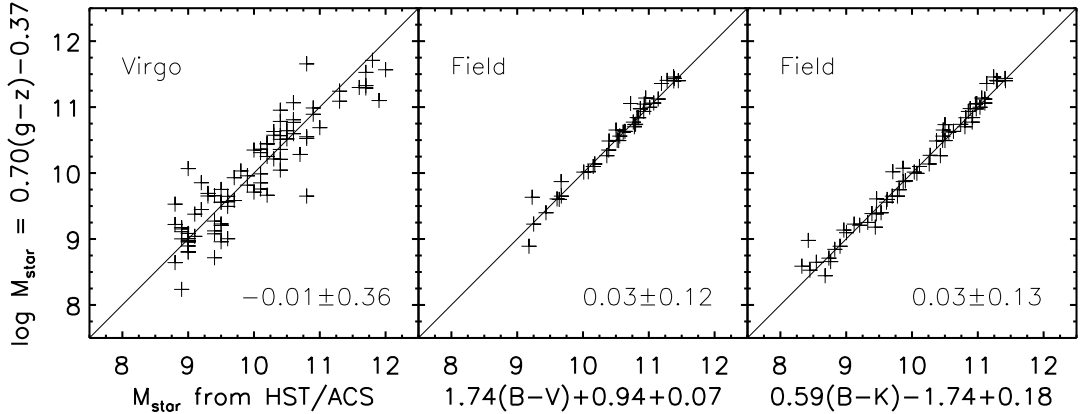} \figcaption{\small
Illustration of methodology used to calculate host galaxy stellar
mass, $M_{\rm star}$. For the Field sample, $M_{\rm star}$ is
calculated from absolute $B$ magnitude and either $g-z$, $B-V$, or
$B-K$ color, in that order of precedence; see $\S$4.1 for
details. After adjustment to the median offset, the different colors
give consistent $M_{\rm star}$.}
\end{figure*}

\section{Global properties of the AMUSE-Field sample}

Here we determine and discuss the global properties of the AMUSE-Field
sample, including stellar and black hole masses as well as
Eddington-scaled \hbox{X-ray} luminosities. We then proceed to
investigate the scaling between the measured nuclear \hbox{X-ray}
luminosity and the host galaxy stellar mass, and we calculate the
residual \hbox{X-ray} luminosity after accounting for this
dependence. Finally, we test against the full Field sample the
properties of the snapshot targets (which comprise a
uniform-sensitivity subset) and of elliptical galaxies (to check for
possible systematic morphological effects).

Table~3 contains the derived properties of the full AMUSE-Field sample
as well as three subsets thereof, composed of: 1) snapshot targets; 2)
elliptical galaxies, and 3) group members. (The influence of group
richness is discussed in more detail in $\S$5.) For each of the above
subsets, along with the full sample, the mean and 25th, 50th, and 75th
percentiles are calculated for the following quantities: $\log{(M_{\rm
star}/M_{\odot})}$, $\log{(M_{\rm BH}/M_{\odot})}$, $\log{(M_{\rm
BH}/M_{\rm star})}$, $M_{\rm B}$, and $\log{L_{\rm X}}$ for both
detections and for all objects. The Kaplan-Meier distribution, which
handles censored data (Kaplan \& Meier 1968), was determined for
\hbox{X-ray} luminosities including upper limits using the survival
analysis package
ASURV\footnote{http://astrostatistics.psu.edu/statcodes/asurv} (e.g.,
Lavalley et al.~1992).\footnote{Note, however, that at least the 25th
percentile values are dominated by upper limits, and the mean values
are biased due to the conversion of a limit to a detection for
calculation purposes, so these values should be taken as roughly
indicative only.} To reduce any potential bias from archival
observations, we conservatively impose a luminosity limit of
$\log{L_{\rm X}}>38.2$ when examining the distribution of $\log{L_{\rm
X}}$ (or quantities derived therefrom) with limits included, and when
carrying out fitting of $L_{\rm X}(M_{\rm star})$; six objects (two
detected) from the full Field sample do not meet this screening
criteria. The adopted methodology and calculation of these quantities
is described in detail in the following sub-sections.

\subsection{Stellar and black hole masses}

The host galaxy stellar mass is calculated from the absolute $B$
magnitude and from optical/IR color. Absolute $B$ magnitudes for the
total galaxy (for these sources, essentially equivalent to the
spheroidal luminosities) are taken from the HyperLeda catalog. These
are based on total $B$ magnitudes reduced to the RC3 system and
corrected for Galactic absorption, and calculated for distances
identical to those used to determine $L_{\rm X}$. We calibrate our
procedure against the stellar masses established for the galaxies in
the AMUSE-Virgo survey, which were determined based on F457W and
F850LP {\it HST}/ACS imaging (G08 $\S$3.3); these filters roughly
correspond to SDSS $g$ and $z$ bands (Fukugita et al.~1996) and
provide nearly equivalent colors. The primary relation used to
calculate stellar mass is \hbox{$M_{\rm star} =
0.70\times(g-z)-0.37+0.4\times(5.45-M_{\rm B})$} (based on Table~7 of
Bell et al.~2003); Figure~3 illustrates that using SDSS $g-z$
model-magnitude colors (corrected for Galactic extinction) gives
results consistent with those obtained from the higher-resolution {\it
HST}/ACS data, with a mean difference of $-0.01\pm0.36$. Where SDSS
coverage is not available, we use instead corrected $B-V$ colors
(taken from HyperLeda) as $M_{\rm star} =
1.74\times(B-V)+0.94+0.4\times(5.45-M_{\rm B})$ (based on Table~7 of
Bell et al.~2003), but here we also add 0.07 which is the empirical
median offset between $M_{\rm star}$ derived from $g-z$ versus $B-V$
colors. No physical meaning is ascribed to this adjustment, which may
merely arise from inconsistent effective apertures; similar median
offsets are found for AMUSE-Virgo galaxies. The agreement between
these methods is good (Figure~3; mean difference $0.03\pm0.12$). Where
neither SDSS coverage nor HyperLeda $B-V$ magnitudes are available, we
use $B-K$ colors (with total $K$ magnitudes taken from the 2MASS
catalog of extended sources) as $M_{\rm star} =
0.59\times(B-K)-1.74+0.4\times(5.45-M_{\rm B})$ (based on Table~1 of
Bell \& de~Jong 2001), but here we also add 0.18 which is the
empirical median offset between $M_{\rm star}$ derived from $g-z$
versus $B-K$ colors. The agreement between the $g-z$ and $B-K$ methods
is also good (Figure~3; mean difference $0.03\pm0.13$). Finally, in
the four cases (PGC 132768, PGC 064718, PGC 740586, and 6dF
J2049400$-$324154) where neither SDSS coverage, nor HyperLeda $B-V$
colors, nor 2MASS extended-source $K$ magnitudes, were available, the
median $g-z=1.16$ is used to estimate $M_{\rm star}$. The \hbox{X-ray}
detection fraction for the Field sample is illustrated versus $M_{\rm
star}$ in Figure~4.

\begin{figure}
\includegraphics[scale=0.5]{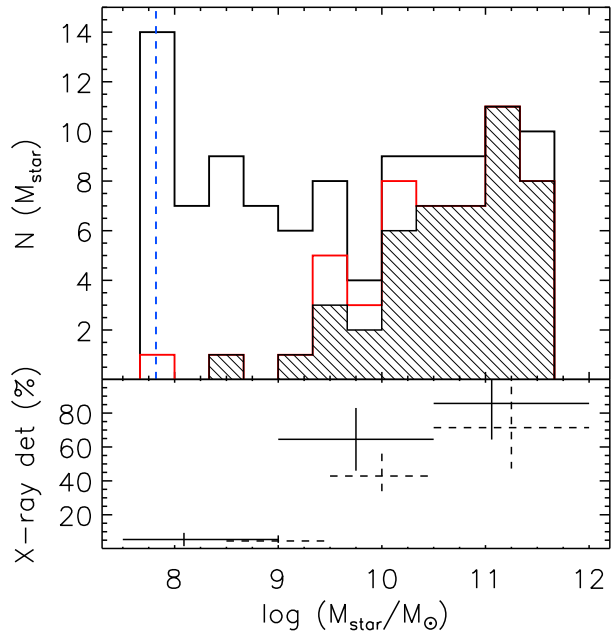} \figcaption{\small The
top panel shows the distribution of stellar mass within the Field
sample. The red histogram shows nuclear X-ray detections, while the
shaded histogram incorporates a statistical correction for potential
LMXB contamination from nuclear star clusters. The bottom panel shows
the X-ray detection fraction (solid crosses) within bins of low,
moderate, and high stellar mass. The increased detection fraction with
increasing stellar mass (e.g., Ho et al. 1997) results here primarily
from an ability to detect X-ray emission to a lower Eddington fraction
in higher-mass objects. A qualitatively similar effect was found for
the Virgo sample (dashed crosses in bottom panel).}
\end{figure}

Black hole masses (\bhm) are estimated from the mass-stellar velocity
dispersion (\hbox{$M_{\rm BH}-\sigma$}) or else from the mass-bulge
luminosity ($M_{\rm BH}-L_{\rm B}$) correlations (e.g., see overview
by Ferrarese \& Ford 2005). These have been calibrated with
independent measures of $M_{\rm BH}$, such as yielded by reverberation
mapping (Woo et al.~2010), enabling straight-forward estimation of
$M_{\rm BH}$. We use the updated relations presented in G{\"u}ltekin
et al.~(2009), specifically those derived for early-type galaxies. As
they find the scatter in the $M_{\rm BH}-\sigma$ relation to be
slightly lower than in the $M_{\rm BH}-L$ relation for early-type
galaxies, we use the former where high-quality measurements of
$\sigma$ are available in the literature. We note that caution is
warranted when applying this correlation to a heterogeneous sample, as
it is not clear that the $M_{\rm BH}-\sigma$ relation holds in
identical form to low masses (Greene et al.~2010) or high luminosities
(Lauer et al.~2007) or for barred galaxies (Graham 2008) or to higher
redshifts (Treu et al.~2007; Woo et al.~2008). Only the first of these
points is potentially relevant for this sample, and in practice there
is only one Field galaxy with $\sigma<25$~km~s$^{-1}$, a measurement
we discard due to concerns as to systematic errors.

Eddington luminosities are calculated from the above-determined black
hole mass as $L_{\rm Edd} = 1.3\times10^{38}(M_{\rm
BH}/M_{\odot})$~erg~s$^{-1}$, and Eddington-scaled \hbox{X-ray}
luminosities, $\log{(L_{\rm X}/L_{\rm Edd})}$, for the AMUSE-Field
sample are plotted in Figure~5 versus both $M_{\rm BH}$ and $M_{\rm
star}$. Values from the AMUSE-Virgo sample are also shown for
comparison. The distributions of $M_{\rm BH}$ and $M_{\rm star}$ are
formally inconsistent ($KS$-test $p=0.01$ and $p<0.001$, respectively)
between the Field and Virgo samples, but qualitatively similar general
trends with Eddington-scaled \hbox{X-ray} luminosities are observed.

{\it It appears that a fixed $L_{\rm X}/L_{\rm Edd}$ value (or range)
across $5<\log{(M_{\rm BH}/M_{\odot})}<9$ does not provide an adequate
match to the observed data, which show a scarcity of large $L_{\rm
X}/L_{\rm Edd}$ values at large $M_{\rm BH}$.}  All objects are highly
sub-Eddington, with Eddington-scaled nuclear X-ray luminosities
ranging from $<10^{-8}$ to $10^{-4}$. Clearly, the limiting
sensitivity of the survey ($\sim 2.5 \times 10^{38}$ \es~for snapshot
targets) acts to restrict the observable Eddington ratio for detected
sources: due to the correlation between black hole mass and host
stellar mass (linked to the underlying dependence of stellar mass on
bulge optical luminosity), nuclear \hbox{X-ray} sources may be
detected down to lower Eddington-scaled luminosities within higher
$M_{\rm star}$ galaxies, and this Eddington-incompleteness effect can
explain the increase in the X-ray detection fraction with increasing
stellar mass seen in Figure~4. This effect was noted by G10 for the
Virgo sample. Despite these observational complications, the relative
weakness of SMBH activity associated with the highest black hole
masses in the Field sample is a qualitatively similar trend to the
``downsizing'' tendency noted by G10.

\begin{figure*}
\centering
\mbox{\subfigure{\includegraphics[scale=0.45]{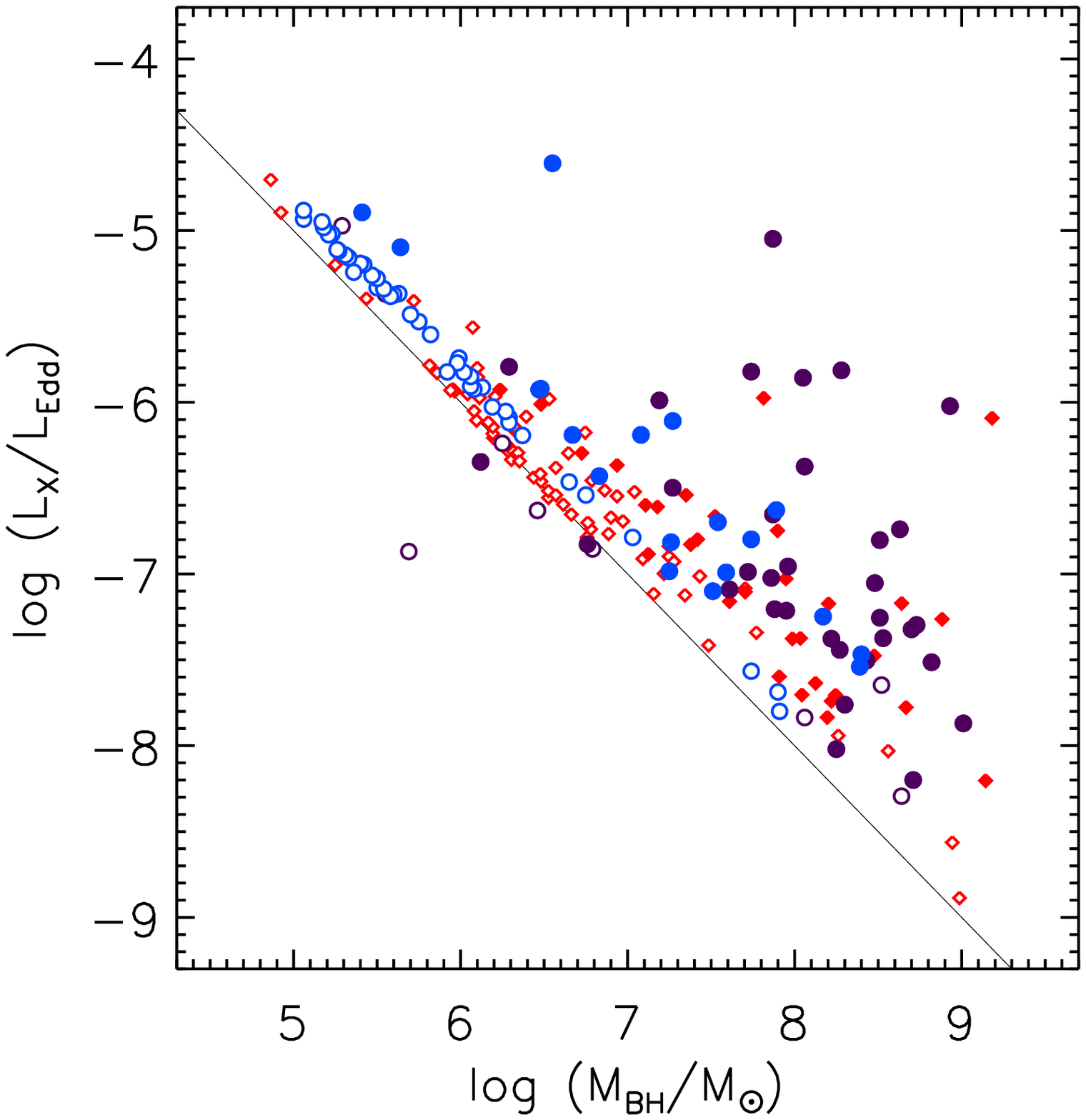}
\quad
\subfigure{\includegraphics[scale=0.45]{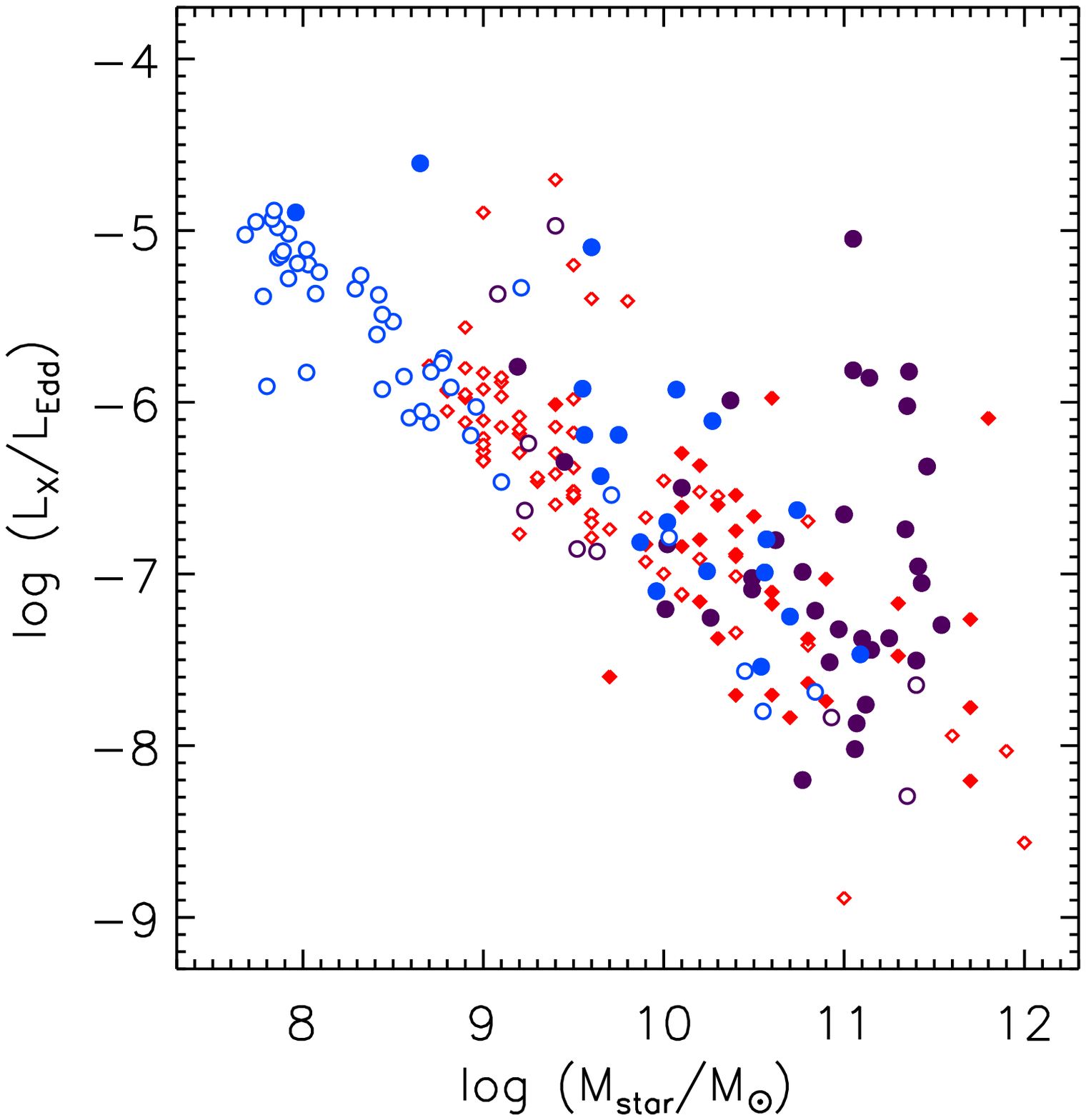}}}}

\figcaption{\small The ratio of X-ray to Eddington luminosity as a
function of black hole mass (left panel) and stellar mass (right
panel). The Field sample is plotted with circles (snapshot/archival
sources are blue/purple) and the Virgo sample is also plotted, with
red diamonds, for comparison. X-ray detections/limits are shown as
filled/open symbols.  The solid black line in the left panel indicates
the sensitivity limit of the Virgo survey, corresponding to the
Eddington limit for a one solar mass black hole. While the
distributions of black hole and stellar mass are formally inconsistent
between the field and Virgo samples, qualitatively similar trends with
Eddington-scaled X-ray luminosity are observed.}
\end{figure*}

\subsection{Nuclear X-ray luminosity as a function of host stellar mass}

In the following we assess the relationship between nuclear
\hbox{X-ray} luminosity and stellar mass within the AMUSE-Field
sample, to investigate how activity in the nucleus is linked to the
large-scale properties of the host galaxy. As $M_{\rm star}$ is
calculated from $M_{\rm B}$ as well as color, this analysis is similar
to examining a luminosity-luminosity correlation and is grounded in
observed quantities. It is reasonable to expect that some factor of
the observed correlation between nuclear X-ray luminosity and stellar
mass is merely reflective of a correlation between nuclear X-ray
luminosity and black hole mass; because $M_{\rm star}$ and $M_{\rm
BH}$ are correlated and, for many of these objects, both based on
$M_{\rm B}$, it is difficult to disentangle their relative influence
upon nuclear \hbox{X-ray} luminosities. Here we restrict out analysis
to nuclear X-ray luminosity considered as a function of host stellar
mass; we defer study of the $L_{\rm X}(M_{\rm BH})$ relation to a
later work (Miller et al.~2012), in which we compare the functional
form of this relation within the AMUSE-Field sample to that found for
the AMUSE-Virgo cluster early-types (G10), while fully accounting for
the multiple methods of determining $M_{\rm BH}$ and for the
inconsistency in $M_{\rm star}$ distributions between the Field and
Virgo samples.

Fitting is carried out using the Bayesian IDL code of Kelly (2007),
which incorporates both uncertainties and censoring, to determine the
best-fit parameters. The functional form investigated is $(\log{L_{\rm
X}}-38.4)=A+B\times(\log{M_{\rm star}}-9.8)$, where the variables are
centered on approximately the median values. Points with $\log{L_{\rm
X}}<38.2$ are excluded to avoid any potential bias from archival
observations, but this has only a minimal effect on the result (e.g.,
the slope is insignificantly flattened by 0.07$\pm$0.14). Errors are
taken to be 0.1 on $\log{L_{\rm X}}$ and 0.1 on $\log{(M_{\rm
star}/M_{\odot})}$, in both cases associated with the uncertainty in
the distance rather than measurement error of the \hbox{X-ray} counts
or the bulge luminosity. Several alternative forms of errors were
investigated, including errors increasing with decreasing
luminosities; in general, larger errors produce a slightly steeper
slope but do not appreciably affect any of the below
conclusions. Three Gaussians are used in the independent variable
mixture modeling, and a minimum of 5000 iterations are performed with
Gibbs sampling. The most likely parameter values are estimated as the
median of 10000 draws from the posterior distribution, with credible
intervals corresponding to 1$\sigma$ errors calculated as the 16th and
84th percentiles. For completeness, we verified that using instead
ASURV with either the expectation-maximization or Buckley-James method
(Buckley \& James 1979) identifies similar trends, with slightly
flatter slopes.

Results are illustrated in Figure~6, with best-fit coefficients listed
in Table~4. For the full AMUSE-Field sample, with all objects
included, the best fit relation is $(\log{L_{\rm
X}}-38.4)=(-0.04\pm0.12)+(0.71\pm0.10)\times(\log{M_{\rm star}}-9.8)$,
with a rather large intrinsic scatter of
\hbox{$\sigma=0.73\pm0.09$}. Consistent results are obtained for the
subsets of snapshot targets, ellipticals, and group members. If only
\hbox{X-ray} detected points are considered, the slope is considerably
flatter, $B=0.35\pm0.12$ for the full sample. Residual \hbox{X-ray}
luminosities may now be calculated as ${\Delta}L_{\rm X,M*} =
\log{L_{\rm X}} - \log{L_{\rm X}}(\log{M_{\rm star}})$, where
$\log{L_{\rm X}}(\log{M_{\rm star}})$ is the best-fit linear relation
for the full Field sample as defined above. ${\Delta}L_{\rm X,M*}$
thus provides a measure of relative \hbox{X-ray} brightness after
accounting for the typical influence of stellar mass.

As $M_{\rm star}$ is derived from absolute $B$ luminosity, the
correlation we find for $L_{\rm X}$ as a function of $M_{\rm star}$ is
qualitatively similar to the tendency noted by Pellegrini (2010) for
$L_{\rm X}$ to increase with $L_{\rm B}$. The slope for the full
AMUSE-Field sample of $B=0.71\pm0.10$ is incompatible with zero
dependence, but also appears to be less than unity (at the $2.9\sigma$
level), indicating that nuclear \hbox{X-ray} luminosity may increase
less rapidly than $M_{\rm star}$: this would imply lower-mass galaxies
are more \hbox{X-ray} luminous per unit stellar mass. While the
general trend in Figure~6 is clear, the observed dispersion in both
$L_{\rm X}$ and ${\Delta}L_{\rm X,M*}$ is larger for $\log{M_{\rm
star}}>10.5 M_{\odot}$, and so it is possible that a more complex
model could provide an improved representation of $L_{\rm X}$ across
the full range of stellar masses.

\subsection{Uniform-sensitivity and morphologically distinct sub-samples}

While the AMUSE-Field sample is by design fairly homogeneous, it does
include both snapshot and archival objects, and both type E and type
E-S0 galaxies. The snapshot sample possesses an approximately constant
\hbox{X-ray} luminosity sensitivity threshold due to the deliberate
choice of exposure times, which is not the case for the archival
sample. Elliptical galaxies are completely spheroidal, whereas
transition objects may include small disk components. It is thus
useful to check the characteristics of these subsets.

{\it Snapshot targets\/}: The archival {\it Chandra\/} coverage
available prior to selection of the snapshot sample understandably
favored brighter galaxies expected to provide a high \hbox{X-ray}
count rate. Within our volume-limited sample, the snapshot targets
therefore tend to be galaxies of lower optical luminosity, and lower
stellar mass. The ratios of black hole to stellar mass [i.e.,
$\log{(M_{\rm BH}/M_{\rm star})}$] are consistent with each other (KS
$p=0.29$) between the snapshot targets and the full Field sample, as
$\log{M_{\rm BH}}$ is also lower for the snapshot targets. (Note that
for a majority of the snapshot targets the $M_{\rm BH}-L$ relation is
used to calculate $M_{\rm BH}$ and so black hole mass is directly
dependent on optical luminosity. The $M_{\rm BH}-L$ relation predicts
slightly higher $M_{\rm BH}$ values at low luminosities; see, e.g.,
discussion in G08.) The snapshot targets tend toward modestly lower
\hbox{X-ray} luminosities where detected, and the difference is
enhanced when upper limits are included (due to the lower detection
fraction for the snapshot targets; $\S$3.1); the snapshot targets have
lower mean and Kaplan-Meier quartile $\log{L_{\rm x}}$ values, and
two-sample tests conducted within ASURV indicate that the snapshot
distribution of $\log{L_{\rm X}}$ is inconsistent with that of the
full Field sample (logrank $p=0.006$, Peto \& Prentice $p=0.003$;
Latta 1981). However, the best-fit $\log{L_{\rm X}}(\log{M_{\rm
star}})$ relation for snapshot targets has an intercept and slope
consistent with that for the full Field sample. Further, the
distribution of residual \hbox{X-ray} luminosities for the snapshot
targets matches that of the full Field sample; the ${\Delta}L_{\rm
X,M*}$ Kaplan-Meier quartiles are $-0.29$, 0.07, 0.37 for the snapshot
targets and $-0.36$, 0.05, 0.38 for the Field sample, and two-sample
tests also support consistency (logrank and Peto \& Prentice
$p>0.5$). This confirms that after accounting for the dependence of
$L_{\rm X}$ upon $M_{\rm star}$, the snapshot targets have
\hbox{X-ray} properties in line with those of the full Field
sample. As the snapshot observations are sensitive to a uniform
limiting luminosity ($\S$2), we further conclude that the full Field
sample is not biased by the wide variety in exposure times of the
archival objects.

\begin{figure}
\includegraphics[scale=0.49]{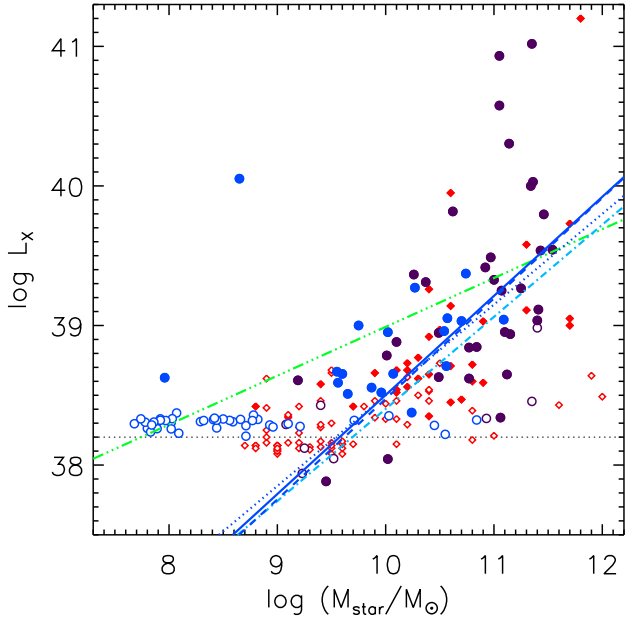}
\figcaption{\small X-ray luminosity as a function of host galaxy
stellar mass. Symbols are as in Figure~5. The solid line is the
best-fit linear relation (to logarithmic quantities, for objects with
$\log{L_{\rm X}}>38.2$) for the full Field sample, while the
dot-dashed, dotted, and dashed lines are for the snapshot, elliptical,
and group member subsets; see $\S$4.2 for details. For reference, the
triple-dot-dashed line is the fit to \hbox{X-ray} detections only. All
fit coefficients are provided in Table~4.}
\end{figure}

{\it Elliptical galaxies\/}: Most (93/103) of the galaxies in the
Field sample are ellipticals, and so it is not surprising that the
properties of these ellipticals are virtually identical to those of
the full Field sample. The detection fraction for ellipticals is 46/93
or 49\%, whereas for non-ellipticals (including three somewhat
irregular spheroids unclassified in HyperLeda) it is 6/10 or 60\% (for
E-S0 galaxies in particular it is 6/7). While the mean or median
$L_{\rm X}$ values for objects are $\sim0.5$ dex greater for the 6
\hbox{X-ray} detected type E-S0 galaxies than for the 46 \hbox{X-ray}
detected type E galaxies, the same holds true for $M_{\rm star}$
values, and so the mean or median ${\Delta}L_{\rm X,M*}$ agree to
within 0.1. In any event the scatter is sufficiently large that both
the $\log{L_{\rm X}}$ and ${\Delta}L_{\rm X,M*}$ distributions for
detected objects are consistent (KS $p>0.5$) between field type E or
type E-S0 galaxies; more quantitative assessment of any minor
differences in nuclear \hbox{X-ray} properties would require a much
larger sample of type E-S0 galaxies. It would also be interesting to
examine a companion sample of S0 galaxies, for which we might expect
the active fraction and the typical nuclear \hbox{X-ray} luminosity to
be somewhat higher than for ellipticals, based on the increasing AGN
fraction toward later morphological types and bluer galaxies found by
Choi et al.~(2009) and attributed by them to a relatively greater gas
supply.

\section{Influence of group richness on nuclear activity}

Although by construction the galaxies in the AMUSE-Field sample do not
reside in clusters (recall from $\S$2 that the coordinate selection
excluded objects in the direction of Fornax and Virgo), some of them
are members of groups, inhabiting regions characterized by galaxy
over-densities intermediate between isolated and cluster environments.
Groups provide a suitable environment for strong interactions and
mergers, since the number density is higher than for the field but the
dispersion in radial velocities of galaxies in groups is smaller than
that found in clusters (recall the dynamical friction in a close
encounter between two galaxies scales inversely with the square of
their relative velocities). Tidal and ram-pressure stripping become
increasingly important effects in richer groups (and clusters). As a
consequence, membership in a group, and the richness of that group,
could have an impact on the interaction history of a galaxy and on its
gas content, both of which may be potentially relevant to the fueling
of the SMBH.  Here we investigate whether group membership or group
properties influence nuclear \hbox{X-ray} activity.

We use the newly available catalog by Makarov \& Karachentsev (2011)
to determine group membership, if any, for each target. There are 8
AMUSE-Field objects not present in this catalog, and so the following
analysis is limited to 95 objects. Following their approach, groups
consist of four or more associated galaxies present within the
zero-velocity surface; 74 (78\%) AMUSE-Field galaxies satisfy this
criteria. The largest group included here is that centered on
NGC~5846, which contains 74 members; the Field sample includes 18 of
these, of which 3 have \hbox{X-ray} detections (the nucleus of
NGC~5846 itself is not significantly detected above the diffuse
emission; see Appendix A). Overall, the detection fraction for group
members is 40/74 or 54\%, while for objects with $n_{\rm group}<4$ (we
will refer to these as ``non-group'' members but it is understood that
they are also obviously not cluster members) it is 11/21 or 52\%;
these rates are consistent with being identical. There are 12
``isolated'' galaxies (with $n_{\rm group}=1$, although these may not
be true void galaxies as in, e.g., Kreckel et al.~2011), of which 5
have \hbox{X-ray} detections, or 42\%.

Restricting consideration to \hbox{X-ray} detected objects, the mean
or median $L_{\rm X}$ values for non-group members are consistent with
those for group members (difference $<0.1$ dex), but as the $M_{\rm
star}$ values of the non-group members tend to be lower by $\simeq$0.6
dex, their mean and median ${\Delta}L_{\rm X,M*}$ values are higher
(by $\simeq$0.4 and $\simeq$0.3, respectively). The scatter is large
enough that the distributions of $\log{L_{\rm X}}$ and ${\Delta}L_{\rm
X,M*}$ are consistent (KS $p>0.2$) between group members or
non-members. The 5 \hbox{X-ray} detected isolated\footnote{We note
that ESO 540$-$014, isolated by these criteria, is an \hbox{X-ray}
bright outlier.} galaxies have even lower $M_{\rm star}$ values than
other non-group members, and consequent higher ${\Delta}L_{\rm X,M*}$
values (overcoming marginally lower $L_{\rm X}$ values); here again
the distributions of $\log{L_{\rm X}}$ and ${\Delta}L_{\rm X,M*}$ are
not formally inconsistent at the $\alpha=0.05$ level (KS $p=0.38$ and
$p=0.06$, respectively; the latter case is perhaps borderline). 

When upper limits are incorporated, similar conclusions
hold. Comparing non-group to group members, the distributions in
$\log{L_{\rm X}}$ or ${\Delta}L_{\rm X,M*}$ are consistent (logrank
and Peto \& Prentice $p>0.4$), despite slightly lower $\log{L_{\rm
X}}$ and slightly higher ${\Delta}L_{\rm X,M*}$ values (by $\sim$0.2
dex) in non-group members. The Kaplan-Meier quartiles are
systematically somewhat lower in $\log{L_{\rm X}}$ for isolated
galaxies compared to group members (0.54, 1.08, 1.61 versus 0.70,
1.62, 2.11), but somewhat greater in ${\Delta}L_{\rm X,M*}$ (0.06,
0.32, 0.47 versus $-0.40$, 0.03, 0.30). The distributions are not
formally inconsistent (logrank and Peto \& Prentice
\hbox{$p>0.1$}). Eddington-scaled \hbox{X-ray} luminosities also
display an apparent progression from lower to higher values moving
from group to non-group to isolated galaxies, but again the
significance is borderline.\footnote{Including upper limits, the mean
$\log{L_{\rm X}/L_{\rm Edd}}$ values for group members, non-group
members, and isolated galaxies are $-7.06\pm0.11$, $-6.62\pm0.20$, and
$-6.26\pm0.19$, with median values of $-7.22$, $-6.98$, $-6.63$;
removing ESO 540$-$014 from the isolated subset gives instead a mean
of $-6.41\pm0.14$ and median of $-6.69$. Two-sample tests for group
members versus non-group members, isolated galaxies, or isolated
galaxies excluding ESO 540$-$014 give logrank $p=0.10$, 0.01, 0.05 and
Peto \& Prentice $p=0.10$, 0.03, 0.12.}

{\it We conclude that there is suggestive but inconclusive evidence
that non-group members, particularly isolated galaxies, tend toward
slightly enhanced values of $\log{L_{\rm X}/L_{\rm Edd}}$ or
${\Delta}L_{\rm X,M*}$ as compared to group members.} If these results
are confirmed, a possible interpretation is that interactions with the
intra-group medium and other group members act to remove gas that
could otherwise become available for SMBH accretion. This may be a
secondary effect; for example, if star formation near the galaxy
center were reduced by environmental interactions, less mass would be
provided to the SMBH from stellar winds.

The nuclear \hbox{X-ray} luminosity, Eddington-scaled \hbox{X-ray}
luminosity, and residual \hbox{X-ray} luminosity are plotted versus
group size and group radial velocity dispersion in Figure~7. We also
include the 100 cluster spheroids from the AMUSE-Virgo sample (G08,
G10), with artificial random scatter added to their abscissa
coordinates for improved visual clarity. The stellar mass of each
galaxy is indicated by the size of the associated marker. While no
strong trends are present and the scatter in \hbox{X-ray} properties
is large, there appears to be a progression from generally higher to
generally lower values of Eddington-scaled and residual \hbox{X-ray}
luminosity moving from isolated to group to cluster environments. As
groups are more hospitable to strong galaxy-galaxy interactions than
are clusters (due to lower galaxy velocity dispersions), a continuous
progression implies that merger history may not be the dominant factor
controlling current nuclear \hbox{X-ray} activity. Instead,
interactions with the intra-group or intra-cluster medium, which
increase in importance as the galaxy velocity dispersions and the
density of the medium increase, may influence low-level SMBH
fueling. (A more quantitative and statistically-robust comparison of
\hbox{X-ray} luminosities in field versus cluster environments is
provided in Miller et al.~2012.)

\begin{figure}
\includegraphics[scale=0.60]{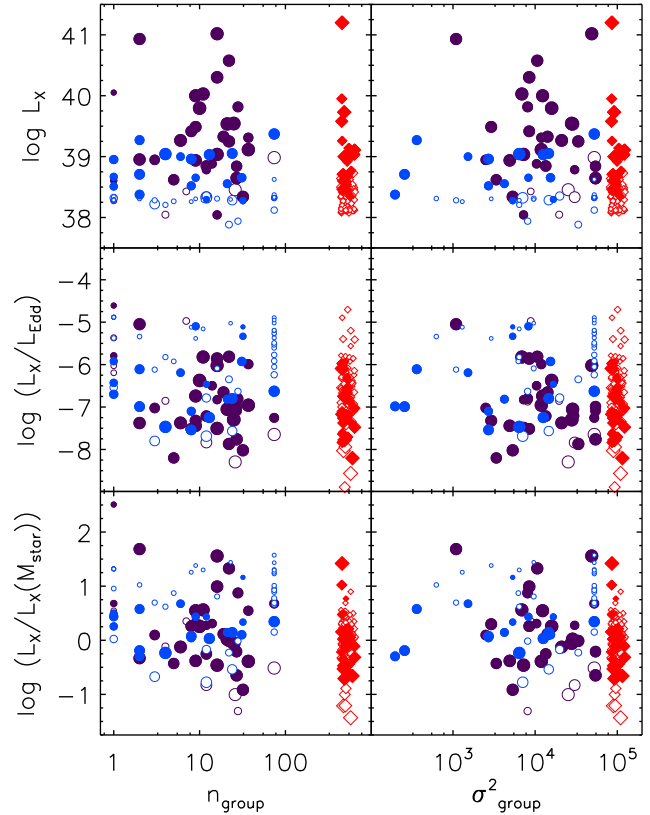} 
\figcaption{\small Nuclear X-ray luminosity (top), Eddington-scaled
\hbox{X-ray} luminosity (middle), and residual \hbox{X-ray} luminosity
(bottom) plotted as a function of group size (left) and group velocity
dispersion (right; note plots truncate lower values of
${\sigma}^{2}_{\rm group}$). Symbols are as in Figure~5 except that
symbol size is here indicative of stellar mass. Artificial $x$--axis
scatter has been added to the Virgo points to improve visual
clarity.}
\end{figure}

Previous works examining the role of environment on \hbox{X-ray}
emission in early-type galaxies found a trend opposite to that
discussed above. A study using {\it ROSAT\/} data carried out by Brown
\& Bregman~(2000) concluded that early-types in richer environments
tended to be more \hbox{X-ray} luminous, with a positive correlation
between $L_{\rm X}/L_{\rm B}$ and galaxy density (but see also
O'Sullivan et al.~2001). They speculated that this could reflect the
influence of the hot intergalactic gas, either through acting to
stifle galactic winds or more directly as an accretion source. We see
no evidence for such a trend in the AMUSE-Field nuclear \hbox{X-ray}
luminosities. This discrepancy is almost certainly due to the much
finer spatial resolution of \cxo, which enables us to disentangle
nuclear, accretion powered \hbox{X-ray} emission from soft diffuse
\hbox{X-ray} emission from gas. In support of this conclusion, we note
that the above {\it ROSAT\/} sample included several galaxies that are
also members of our archival sample (NGC 0720, 1407, 4125, 4278, 4494,
4636, 4697, 5322, 5846) and that the X-ray luminosities measured by
{\it ROSAT\/} are generally 1--2 orders of magnitude larger than those
measured by \cxo.

\section{Summary}

This work presents first results from the AMUSE-Field survey, a Large
{\it Chandra\/} program designed to characterize low-level SMBH
activity within nearby field spheroids. The Field sample contains 103
early-type non-cluster galaxies within 30~Mpc, selected entirely based
on their optical properties and spanning a wide range in absolute
magnitude and stellar mass. For 61 objects we obtained new ACIS-S
observations to a uniform limiting luminosity of
$2.5\times10^{38}$~erg~s$^{-1}$, and for the remaining 42 objects we
make use of high-quality archival coverage. The primary results of
this paper are as follows:

1. The detection fraction of nuclear \hbox{X-ray} sources is 52/103. A
   handful of these may be contaminated by LMXBs; we estimate the
   fraction of accreting SMBHs is $45\pm7$\%. This provides a firm
   lower limit on the SMBH occupation fraction within field spheroids.

2. The ratio of \hbox{X-ray} to Eddington luminosities within the
   Field sample is in all cases low, from $\sim10^{-4}-10^{-8}$. This
   range is similar to that found in the AMUSE-Virgo study of cluster
   early-types (G10). Independent of environment, SMBH accretion
   within local early-types is generally highly sub-Eddington. 

3. We quantify the correlation between nuclear \hbox{X-ray} luminosity
   and host stellar mass, finding a best-fit relation of $(\log{L_{\rm
   X}}-38.4)=(-0.04\pm0.12)+(0.71\pm0.10)\times(\log{M_{\rm
   star}}-9.8)$. The calculated slope is incompatible with zero
   dependence and marginally indicative ($2.9\sigma$) of greater
   $L_{\rm X}$ per unit $M_{\rm star}$ in lower-mass galaxies.

4. After scaling by black hole mass or accounting for the influence of
   stellar mass, typical values of $\log{L_{\rm X}/L_{\rm Edd}}$ or
   ${\Delta}L_{\rm X,M*}$ of group members may be slightly lower than
   for isolated galaxies, and this newly-identified trend appears to
   continue to cluster early-types. If confirmed, a progressive
   decrease in relative \hbox{X-ray} luminosity with increasing
   richness mandates an environmental modulation of SMBH fueling.

Several complementary projects that build on this dataset are
underway. A direct comparison of the nuclear \hbox{X-ray} luminosity
as a function of black hole mass within field and cluster galaxies,
controlling for stellar mass, is discussed in Miller et al.~(2012). An
additional later work will present the catalog of off-nuclear
\hbox{X-ray} sources within the field sample, and compare the
distribution of ULXs to that found for Virgo. We are also currently
obtaining high-resolution {\it HST\/}/ACS imaging of Field galaxies
with \hbox{X-ray} detections, to check for and characterize any
nuclear star clusters.

\acknowledgments

We thank Kayhan G{\"u}ltekin for sharing his {\it Chandra\/} data in
advance of publication, we thank him and Marta Volonteri for useful
conversations, and we thank an anonymous referee for constructive
suggestions that improved this paper. T.~T.~acknowledges support from
the NSF through CAREER award NSF-0642621, and from the Packard
Foundation through a Packard Research
Fellowship. J.~H.~W.~acknowledges support by the Basic Science
Research Program through the National Research Foundation of Korea
funded by the Ministry of Education, Science and Technology
(2010-0021558). Support for this work was provided by NASA through
{\it Chandra\/} Awards Number 11620915 (AMUSE-Field) and 08900784
(AMUSE-Virgo) issued by the {\it Chandra\/} \hbox{X-ray} Observatory
Center. This research has made use of the NASA/IPAC Extragalactic
Database (NED) which is operated by the Jet Propulsion Laboratory,
California Institute of Technology, under contract with NASA. We
acknowledge the usage of the HyperLeda database.

\appendix

\section{Notes on selected archival $L_{\rm X}$ measurements}

{\bf NGC 5846}: The \hbox{X-ray} morphology of this galaxy is complex,
with a cavity to the Northeast bounded by a ring, with similar
structure also seen in H$\alpha$ (Trinchieri \& Goudfrooij~2002). The
optical center of the galaxy is near the radio core (Dunn et al.~2010)
and corresponds to object 12 in the catalog of Trinchieri \&
Goudfrooij~(2002). Filho et al.~(2004) find $L_{\rm
X}=5.6\times10^{38}$~erg~s$^{-1}$ (0.5--10~keV). While the nucleus is
identified by {\it wavdetect\/} in the full-band image, it is
difficult to estimate the local (gas) background, and we do not find a
significant detection in the hard-band image ($L_{\rm
X}<9.5\times10^{38}$~erg~s$^{-1}$). This is the primary galaxy of the
largest group within the Field sample.

{\bf NGC 3923}: Boroson et al.~(2011) do not find an AGN component to
be detectable in this galaxy, and we agree with this assessment. While
{\it wavdetect\/} finds a potential nuclear source in the full-band
image, the enclosing ellipse is substantially larger than the PSF, and
the hard-band image shows complex extended emission, likely from
several unresolved sources.

{\bf IC 1459}: Fabbiano et al.~(2003) find that the nuclear emission
of this radio galaxy is well-fit with an unabsorbed power-law with
$\Gamma$=1.9, and their $L_{\rm X}$ is consistent with our value of
10$^{41}$~erg~s$^{-1}$; this source has the greatest \hbox{X-ray}
luminosity in the Field sample.

{\bf NGC 4636}: Loewenstein et al.~(2001) describe a ``near-nuclear''
source that we find to be coincident with the galactic center (offset
$\sim$0.4$''$; our applied astrometry correction of 0.1$''$ is less
than their 0.9$''$) and this source matches the location of the radio
core (Dunn et al.~2010). As described in Dunn et al.~(2010), this
galaxy is 2.6~Mpc from M87 and shows disturbed \hbox{X-ray}
morphology.

{\bf NGC 1332}: Humphrey \& Buote (2004) do not find evidence for a
point-like central source, and give a limit of $L_{\rm
X}<8.4\times10^{38}$~erg~s$^{-1}$. We measure a significant detection
at the peak of the hard-band emission, which is concentrated at the
nucleus.

{\bf NGC 5576} and {\bf NGC 2778}: G\"ultekin et al.~(2011) find
somewhat greater (unabsorbed) $L_{\rm X}$ values for these sources,
although due to limited counts the uncertainties are necessarily
large. Their best-fit spectral models have steep power-law indices
($\Gamma\sim3.5$ and 4.6, respectively) and some intrinsic absorption,
which differs significantly from the generic model we use to convert
counts to fluxes.

{\bf NGC 3384}: From an early short exposure, Boroson et al.~(2011)
quote a loose upper limit to the AGN component, whereas Zhang et
al.~(2011) do find an \hbox{X-ray} nucleus with $L_{\rm X}$ matching
our value of $6.3\times10^{38}$~erg~s$^{-1}$; our measurement is based
on a subsequent deeper exposure as detailed in G\"ultekin et
al.~(2011), who also find a similar $L_{\rm X}$.

{\bf NGC 3115}: Wong et al.~(2011) describe the central emission as a
``plateau'' and give only an upper limit for the \hbox{X-ray}
nucleus. Our $L_{\rm X}$ value is consistent with that given by
Boroson et al.~(2011), but it does appear possible that this is
somewhat ($\sim$50\%, or 0.2 dex) overestimated due at least one
additional source crowding the nucleus.

{\bf NGC 4278}: Brassington et al.~(2009) analyze six deep exposures
(combined 458~ks) of this galaxy; the LINER nucleus is object 117 in
their catalog. They find $L_{\rm X}=2.0\times10^{40}$~erg~s$^{-1}$ (as
do Boroson et al.~2011), which is $\sim$0.5 dex greater than our
value; 0.3 dex of the discrepancy is due to their greater distance
(likely more accurate than our radial-velocity value from HyperLeda in
this instance), and the remainder may reflect intrinsic variability:
for the ObsID 7081 we use, Brassington et al.~(2009) find $L_{\rm X}$
lower by 0.2 dex than their overall exposure-weighted mean.

{\bf NGC 3377}: The central source in this galaxy connects to an
Eastward ``bridge'' of emission, which includes at least one
additional source. Boroson et al.~(2011) do not find an AGN component
to be detectable, but Zhang et al.~(2009) list $L_{\rm
X}=3.9\times10^{38}$~erg~s$^{-1}$ (0.3--8~keV), in good agreement with
our value.

{\bf NGC 4627}: This nearby ($\sim$10.4 Mpc) galaxy is serendipitously
present off-axis in a deep exposure targeting NGC 4631, but we do not
detect it in our aperture photometry, and Zhang et al.~(2009) also
give only an upper limit. This source has the most stringent limit on
\hbox{X-ray} luminosity ($L_{\rm X}<8.5\times10^{36}$~erg~s$^{-1}$)
within the Field sample.

\clearpage

\LongTables

\begin{deluxetable}{p{100pt}rrrrrrrrrrr}
\tablecaption{Optical Properties}
\tablewidth{17.5cm}

\tablehead{ \colhead{Name} & \colhead{Type} & \colhead{RA} & \colhead{Dec} &
 \colhead{$d$} & \colhead{$B_{\rm T}$} & \colhead{$M_{\rm B}$} &
 \colhead{$\sigma$} & \colhead{$M_{\rm star}$} & \colhead{$M_{\rm
 BH}$} & \colhead{$n_{\rm group}$} & \colhead{${\sigma}_{\rm group}$}
 \\[+3pt] & & \colhead{(deg)} & \colhead{(deg)} & \colhead{(Mpc)} &
 \colhead{(mag)} & \colhead{(mag)} & \colhead{(km~s$^{-1}$)} &
 \colhead{($M_{\odot}$)} & \colhead{($M_{\odot}$)} & &
 \colhead{(km~s$^{-1}$)}\\[-3pt]}

\startdata

NGC1407 & E &  55.049583 & -18.580278 &  23.2 &  10.4 & -21.5 &          272 & 11.5 &  8.7 &       25 &          167 \\ 
NGC5322 & E & 207.313750 &  60.190556 &  29.4 &  11.0 & -21.4 &          233 & 11.4 &  8.5 &       21 &          169 \\ 
NGC4125 & E & 182.025000 &  65.174167 &  22.9 &  10.5 & -21.3 &          227 & 11.4 &  8.4 &       16 &           85 \\ 
NGC5846 & E & 226.622083 &   1.605556 &  26.2 &  10.8 & -21.3 &          239 & 11.4 &  8.5 &       74 &          229 \\ 
NGC2768 & E & 137.906250 &  60.037222 &  23.1 &  10.6 & -21.2 &          181 & 11.5 &  8.1 &       10 &          126 \\ 
NGC3923 & E & 177.757083 & -28.806111 &  20.4 &  10.4 & -21.1 &          257 & 11.4 &  8.6 &       26 &          159 \\ 
NGC4697 & E & 192.149583 &  -5.800833 &  17.8 &  10.1 & -21.1 &          171 & 11.4 &  8.0 &       37 &          109 \\ 
NGC4494 & E & 187.850417 &  25.775278 &  20.9 &  10.6 & -21.0 &          150 & 11.4 &  7.7 &       11 &           83 \\ 
IC1459 & E & 344.294167 & -36.462222 &  23.2 &  10.9 & -21.0 &          306 & 11.4 &  8.9 &       16 &          221 \\ 
NGC5077 & E & 199.882083 & -12.656944 &  40.2 &  12.1 & -21.0 &          256 & 11.3 &  8.6 &        9 &          111 \\ 
NGC0821 & E &  32.087917 &  10.995000 &  25.0 &  11.2 & -20.7 &          200 & 11.1 &  8.2 &        2 &  \nodata  \\ 
NGC0720 & E &  28.252083 & -13.738611 &  22.8 &  11.1 & -20.7 &          241 & 11.2 &  8.5 &        6 &          145 \\ 
NGC3585 & E & 168.321250 & -26.754722 &  17.8 &  10.5 & -20.7 &          206 & 11.1 &  8.3 &        9 &           70 \\ 
NGC3610 & E & 169.605417 &  58.786389 &  27.8 &  11.6 & -20.7 &          162 & 11.0 &  7.9 &       19 &          119 \\ 
NGC3379 & E & 161.956667 &  12.581667 &  13.6 &  10.1 & -20.5 &          209 & 11.1 &  8.3 &       27 &          233 \\ 
NGC4636 & E & 190.707500 &   2.687778 &  14.5 &  10.3 & -20.5 &          203 & 11.1 &  8.2 &       32 &           73 \\ 
NGC7507 & E & 348.031667 & -28.539722 &  21.0 &  11.1 & -20.5 &          222 & 11.1 &  8.4 &        4 &           80 \\ 
NGC1332 & E-SO &  51.572083 & -21.335278 &  20.0 &  11.0 & -20.5 &          321 & 11.1 &  9.0 &       22 &          183 \\ 
NGC4036 & E-SO & 180.361667 &  61.895833 &  23.2 &  11.4 & -20.4 &          181 & 11.1 &  8.1 &       16 &           92 \\ 
NGC3640 & E & 170.278750 &   3.234722 &  19.0 &  11.1 & -20.3 &          181 & 10.9 &  8.1 &       12 &          174 \\ 
NGC1052 & E &  40.270000 &  -8.255833 &  19.7 &  11.3 & -20.1 &          207 & 11.1 &  8.3 &       22 &          103 \\ 
NGC5576 & E & 215.265417 &   3.271111 &  22.7 &  11.6 & -20.1 &          171 & 10.8 &  7.9 &       12 &          116 \\ 
NGC4291 & E & 185.075833 &  75.370833 &  28.7 &  12.2 & -20.1 &          285 & 10.9 &  8.8 &        8 &           92 \\ 
NGC5838 & E-SO & 226.359583 &   2.099444 &  20.7 &  11.5 & -20.0 &          266 & 11.0 &  8.7 &        9 &           54 \\ 
NGC5638 & E & 217.418333 &   3.233333 &  24.8 &  12.0 & -20.0 &          165 & 10.8 &  7.9 &       12 &           84 \\ 
NGC5831 & E & 226.029167 &   1.220000 &  24.9 &  12.1 & -19.8 &          165 & 10.7 &  7.9 &       74 &          229 \\ 
NGC3384 & E-SO & 162.070417 &  12.629167 &  13.1 &  10.8 & -19.8 &          148 & 10.8 &  7.7 &       27 &          233 \\ 
NGC3115 & E-SO & 151.308333 &  -7.718611 &   8.4 &   9.9 & -19.8 &          268 & 10.8 &  8.7 &        5 &           58 \\ 
NGC3193 & E & 154.603750 &  21.893889 &  20.8 &  11.9 & -19.7 &          194 & 10.7 &  8.2 &       13 &          112 \\ 
NGC4203 & E-SO & 183.771250 &  33.197222 &  18.1 &  11.7 & -19.6 &          162 & 11.1 &  7.9 &        2 &  \nodata  \\ 
NGC1439 & E &  56.208333 & -21.920556 &  21.3 &  12.1 & -19.5 &          150 & 10.6 &  7.7 &       24 &          121 \\ 
NGC5582 & E & 215.179583 &  39.693611 &  23.9 &  12.4 & -19.5 &          137 & 10.6 &  7.6 &        2 &  \nodata  \\ 
NGC1340 & E &  52.082083 & -31.068056 &  12.8 &  11.1 & -19.4 &          166 & 10.6 &  7.9 &        3 &  \nodata  \\ 
NGC4278 & E & 185.028333 &  29.280833 &  11.3 &  10.9 & -19.4 &          237 & 10.6 &  8.5 &       28 &           90 \\ 
NGC4742 & E & 192.950000 & -10.454722 &  17.9 &  11.9 & -19.4 &          108 & 10.4 &  7.2 &       37 &          109 \\ 
NGC2778 & E & 138.101667 &  35.027500 &  30.9 &  13.2 & -19.2 &          162 & 10.5 &  7.9 &        3 &  \nodata  \\ 
NGC4648 & E & 190.435000 &  74.420833 &  25.7 &  12.8 & -19.2 &          221 & 10.5 &  8.4 &        8 &           52 \\ 
NGC3377 & E & 161.926250 &  13.985833 &  10.6 &  11.0 & -19.2 &          139 & 10.5 &  7.6 &       27 &          233 \\ 
NGC1426 & E &  55.704583 & -22.108333 &  18.0 &  12.2 & -19.1 &          151 & 10.4 &  7.7 &       24 &          121 \\ 
NGC1172 & E &  45.400000 & -14.836667 &  19.9 &  12.4 & -19.0 &          112 & 10.2 &  7.2 &        2 &  \nodata  \\ 
NGC6017 & E & 239.314167 &   5.998333 &  27.3 &  13.5 & -18.6 &          113 & 10.3 &  7.3 &        2 &  \nodata  \\ 
NGC5845 & E & 226.503333 &   1.633889 &  21.9 &  13.2 & -18.5 &          238 & 10.3 &  8.5 &       74 &          229 \\ 
NGC3457 & E & 163.702500 &  17.621111 &  17.7 &  12.8 & -18.4 &           71 & 10.1 &  6.5 &       31 &          124 \\ 
ESO576-076 & E & 202.678750 & -22.421111 &  23.0 &  13.5 & -18.3 &           98 & 10.0 &  7.0 &       12 &          141 \\ 
NGC4283 & E & 185.086667 &  29.310833 &  16.4 &  12.9 & -18.2 &          114 & 10.1 &  7.3 &       14 &          102 \\ 
NGC3928 & E & 177.948333 &  48.683056 &  17.2 &  13.1 & -18.1 &          113 &  9.9 &  7.3 &       21 &           65 \\ 
NGC3641 & E & 170.286667 &   3.194722 &  25.5 &  13.9 & -18.1 &          163 & 10.0 &  7.9 &       12 &          174 \\ 
NGC4121 & E & 181.985833 &  65.113889 &  24.0 &  13.8 & -18.1 &           84 & 10.0 &  6.8 &       16 &           85 \\ 
UGC07767 & E & 188.885000 &  73.674722 &  23.2 &  13.8 & -18.1 &          131 & 10.0 &  7.5 &        8 &           52 \\ 
NGC3265 & E & 157.778333 &  28.796667 &  22.4 &  13.9 & -17.8 &  \nodata  &  9.8 &  7.1 &        6 &           39 \\ 
IC1729 & E &  26.980417 & -26.892222 &  17.5 &  13.4 & -17.8 &          133 & 10.0 &  7.5 &        1 &  \nodata  \\ 
NGC2970 & E & 145.879583 &  31.976944 &  25.2 &  14.3 & -17.7 &           43 &  9.6 &  5.6 &        9 &           91 \\ 
UGC05955 & E & 163.017917 &  71.773056 &  21.7 &  14.0 & -17.6 &           83 &  9.7 &  6.8 &        1 &  \nodata  \\ 
NGC3073 & E-SO & 150.217083 &  55.618889 &  20.3 &  14.0 & -17.5 &           35 &  9.4 &  5.3 &        7 &           94 \\ 
NGC3522 & E & 166.668750 &  20.085556 &  18.8 &  14.0 & -17.4 &           87 &  9.6 &  6.8 &        1 &  \nodata  \\ 
NGC1097A & E &  41.541250 & -30.228056 &  16.2 &  13.9 & -17.2 &  \nodata  &  9.5 &  6.8 &        4 &          140 \\ 
NGC4627 & E & 190.498750 &  32.573611 &  10.4 &  12.9 & -17.1 &           44 &  9.6 &  5.7 &       28 &           90 \\ 
NGC1370 & E &  53.810833 & -20.373611 &  12.6 &  13.5 & -17.0 &           71 &  9.6 &  6.5 &        1 &  \nodata  \\ 
NGC0855 & E &  33.514583 &  27.877222 &   9.7 &  13.0 & -17.0 &           63 &  9.2 &  6.3 &        1 &  \nodata  \\ 
NGC7077 & E & 322.498333 &   2.414167 &  17.1 &  14.2 & -16.9 &           41 &  9.1 &  5.6 &        1 &  \nodata  \\ 
PGC056821 & E & 240.697917 &  19.787222 &  26.2 &  15.2 & -16.9 &  \nodata  &  9.6 &  6.7 &  \nodata  &  \nodata  \\ 
IC0225 & E &  36.617917 &   1.160556 &  21.1 &  14.7 & -16.9 &  \nodata  &  9.1 &  6.7 &       12 &          129 \\ 
NGC1331 & E &  51.617917 & -21.355278 &  14.8 &  14.1 & -16.7 &           57 &  9.4 &  6.1 &       22 &          183 \\ 
ESO540-014 & E &  10.298750 & -21.131667 &  21.8 &  15.0 & -16.6 &  \nodata  &  8.6 &  6.6 &        1 &  \nodata  \\ 
PGC042748 & E & 190.734583 &   3.676667 &  14.6 &  14.5 & -16.3 &           39 &  9.2 &  5.5 &       32 &           73 \\ 
NGC4308 & E & 185.487083 &  30.074444 &  11.3 &  14.0 & -16.2 &           70 &  9.2 &  6.5 &       28 &           90 \\ 
NGC5099 & E & 200.331667 & -13.042500 &  18.5 &  15.1 & -16.2 &  \nodata  &  8.9 &  6.4 &        1 &  \nodata  \\ 
PGC042173 & E & 189.451250 &  -1.344722 &  22.3 &  15.7 & -16.1 &  \nodata  &  8.6 &  6.3 &       16 &           98 \\ 
PGC028305 & E & 147.545833 &  28.013056 &  22.2 &  15.7 & -16.0 &  \nodata  &  8.7 &  6.3 &        3 &  \nodata  \\ 
PGC1242097 & E & 224.692083 &   2.969167 &  26.8 &  16.1 & -16.0 &  \nodata  &  8.7 &  6.3 &       74 &          229 \\ 
PGC3119319 & E & 226.642917 &   1.558889 &  22.9 &  15.8 & -16.0 &  \nodata  &  9.2 &  6.2 &       74 &          229 \\ 
PGC132768 & ? &   5.767500 & -27.926944 &  19.8 &  15.6 & -15.8 &  \nodata  &  9.0 &  6.2 &        1 &  \nodata  \\ 
PGC1210284 & E & 227.312500 &   1.921389 &  26.1 &  16.4 & -15.7 &  \nodata  &  8.8 &  6.1 &       74 &          229 \\ 
PGC030133 & E & 154.756250 &  21.283611 &  16.7 &  15.5 & -15.6 &  \nodata  &  8.4 &  6.1 &        5 &           79 \\ 
PGC085239 & E & 338.055417 & -41.169444 &  20.2 &  16.0 & -15.5 &  \nodata  &  7.8 &  6.1 &        1 &  \nodata  \\ 
PGC042737 & ? & 190.711667 &  12.308611 &  25.7 &  16.5 & -15.5 &  \nodata  &  8.6 &  6.1 &  \nodata  &  \nodata  \\ 
PGC3097911 & E &  40.356250 &  -8.126944 &  18.5 &  15.9 & -15.5 &  \nodata  &  8.0 &  6.0 &  \nodata  &  \nodata  \\ 
PGC064718 & E & 306.890833 & -55.090278 &  11.9 &  15.0 & -15.4 &  \nodata  &  8.8 &  6.0 &  \nodata  &  \nodata  \\ 
PGC1209872 & E & 226.460833 &   1.908333 &  25.9 &  16.7 & -15.4 &  \nodata  &  8.8 &  6.0 &       74 &          229 \\ 
PGC740586 & E & 347.945000 & -28.529167 &  18.4 &  16.1 & -15.2 &  \nodata  &  8.7 &  5.9 &        4 &           80 \\ 
PGC1202458 & E & 227.755417 &   1.680556 &  25.7 &  17.0 & -15.0 &  \nodata  &  8.4 &  5.8 &       74 &          229 \\ 
PGC1216386 & E & 226.102917 &   2.114722 &  26.3 &  17.2 & -14.9 &  \nodata  &  8.5 &  5.8 &       74 &          229 \\ 
PGC1230503 & E & 225.934583 &   2.552222 &  26.5 &  17.4 & -14.7 &  \nodata  &  8.4 &  5.7 &       74 &          229 \\ 
SDSSJ150907.83+004329.7 & E & 227.282917 &   0.725000 &  25.0 &  17.4 & -14.6 &  \nodata  &  8.1 &  5.6 &       74 &          229 \\ 
6dFJ2049400-324154 & E & 312.416667 & -32.698056 &  23.3 &  17.3 & -14.5 &  \nodata  &  8.4 &  5.6 &        1 &  \nodata  \\ 
SDSSJ150033.02+021349.1 & E & 225.137500 &   2.230278 &  19.5 &  17.0 & -14.4 &  \nodata  &  7.8 &  5.6 &        9 &           54 \\ 
SDSSJ145828.64+013234.6 & E & 224.619167 &   1.543056 &  22.5 &  17.4 & -14.4 &  \nodata  &  8.3 &  5.5 &        2 &  \nodata  \\ 
SDSSJ150100.85+010049.8 & E & 225.253750 &   1.013889 &  26.2 &  17.8 & -14.3 &  \nodata  &  7.9 &  5.5 &       74 &          229 \\ 
SDSSJ150812.35+012959.7 & E & 227.051667 &   1.499722 &  24.4 &  17.7 & -14.2 &  \nodata  &  8.3 &  5.5 &       74 &          229 \\ 
PGC1179083 & E & 226.099167 &   0.918333 &  24.8 &  17.9 & -14.1 &  \nodata  &  8.0 &  5.4 &       74 &          229 \\ 
PGC1206166 & E & 227.094583 &   1.798611 &  25.1 &  17.9 & -14.1 &  \nodata  &  8.0 &  5.4 &       74 &          229 \\ 
SDSSJ150233.03+015608.3 & E & 225.637500 &   1.935556 &  24.7 &  17.9 & -14.1 &  \nodata  &  8.0 &  5.4 &       74 &          229 \\ 
PGC042724 & E & 190.689167 &   3.430556 &  10.1 &  16.0 & -14.0 &  \nodata  &  8.1 &  5.4 &  \nodata  &  \nodata  \\ 
PGC135659 & E &  40.794167 &  -0.262778 &  12.8 &  16.6 & -13.9 &  \nodata  &  7.9 &  5.3 &       11 &           80 \\ 
PGC135829 & E & 202.891667 &   2.188611 &  20.2 &  17.7 & -13.9 &  \nodata  &  7.9 &  5.3 &  \nodata  &  \nodata  \\ 
PGC135818 & E & 195.934167 &   2.039722 &  14.6 &  17.1 & -13.8 &  \nodata  &  7.9 &  5.3 &        8 &           36 \\ 
PGC042596 & ? & 190.438333 &   4.006667 &  12.3 &  16.7 & -13.8 &  \nodata  &  8.0 &  5.3 &       32 &           73 \\ 
PGC1223766 & E & 224.670417 &   2.339722 &  23.9 &  18.2 & -13.7 &  \nodata  &  7.9 &  5.2 &       74 &          229 \\ 
PGC043421 & E & 192.530417 &   2.248056 &  15.9 &  17.4 & -13.6 &  \nodata  &  7.7 &  5.2 &       23 &           98 \\ 
SDSSJ145944.77+020752.1 & E & 224.936667 &   2.131111 &  22.0 &  18.2 & -13.6 &  \nodata  &  7.9 &  5.2 &  \nodata  &  \nodata  \\ 
PGC1192611 & E & 225.617083 &   1.364167 &  22.8 &  18.2 & -13.5 &  \nodata  &  7.7 &  5.2 &       74 &          229 \\ 
PGC042454 & E & 190.107917 &   4.050278 &  12.1 &  17.1 & -13.3 &  \nodata  &  7.8 &  5.1 &  \nodata  &  \nodata  \\ 
PGC1217593 & E & 227.005833 &   2.151111 &  16.5 &  17.8 & -13.3 &  \nodata  &  7.8 &  5.1 &        1 &  \nodata  \\[-17pt]

\enddata

\tablecomments{The columns Name, Type, RA, Dec, $d$, $B_{\rm T}$,
$M_{\rm B}$, and $\sigma$ are taken directly or indirectly from the
HyperLeda database. Stellar and black hole masses $M_{\rm star}$ and
$M_{\rm BH}$ are given here as logarithm quantities and calculated as
described in $\S$4.1. Group size and velocity dispersions are from
Makarov \& Karachentsev (2011). Rows are ordered by absolute $B$
magnitude.}

\end{deluxetable}

\clearpage

\begin{deluxetable}{p{100pt}rrrrrrrr}
\tablecaption{X-ray Properties}
\tabletypesize{\scriptsize}
\tablewidth{14.2cm}

\tablehead{ \colhead{Name} & \colhead{ObsID} & 
 \colhead{Band} & \colhead{${\theta}$} & \colhead{Exp} & \colhead{Counts} &
 \colhead{Rate} & \colhead{$\log{L_{\rm X}}$} & \colhead{Ref\tablenotemark{a}} \\[+3pt]
 & & (keV) & ($'$) & (ks) & & \colhead{(ks$^{-1}$)} & \colhead{(erg~s$^{-1}$)} & \\[-3pt]}

\startdata

NGC1407                 &   791 &   2-7 &    0.0 &   35.4 &      44.1 &    1.2 & 39.5    & 1  \\ 
NGC5322                 &  6787 &   2-7 &    0.0 &   13.8 &       9.0 &    0.7 & 39.5    & 2  \\ 
NGC4125                 &  2071 &   2-7 &    0.1 &   61.6 &      21.9 &    0.4 & 39.0    & 3  \\ 
NGC5846                 &   788 &   2-7 &    0.1 &   23.1 &       5.4 & $<$0.3 & $<$39.0 & 4  \\ 
NGC2768                 &  9528 &   2-7 &    0.0 &   64.1 &     140.3 &    2.4 & 39.8    & 3  \\ 
NGC3923                 &  9507 &   2-7 &    0.1 &   80.6 &      23.0 & $<$0.1 & $<$38.5 & 5  \\ 
NGC4697                 &  4730 &   2-7 &    0.0 &   38.2 &      26.3 &    0.8 & 39.1    & 6  \\ 
NGC4494                 &  2079 &   2-7 &    0.1 &   15.0 &      61.7 &    4.6 & 40.0    & 7  \\ 
IC1459                  &  2196 &   2-7 &    0.0 &   52.7 &    1706.5 &   36.0 & 41.0    & 8  \\ 
NGC5077                 & 11780 &   2-7 &    0.0 &   28.8 &      31.7 &    1.2 & 40.0    & 9  \\ 
NGC0821                 &  6313 &   2-7 &    0.0 &   49.3 &      11.7 &    0.3 & 39.0    & 3  \\ 
NGC0720                 &  7372 &   2-7 &    0.0 &   47.7 &      29.3 &    0.7 & 39.3    & 10 \\ 
NGC3585                 &  9506 &   2-7 &    0.1 &   59.2 &      30.7 &    0.6 & 38.9    & 8  \\ 
NGC3610                 &  7141 & 0.3-7 &    0.0 &    4.9 &      15.1 &    3.2 & 39.3    & 11 \\ 
NGC3379                 &  7073 &   2-7 &    0.0 &   81.3 &      37.2 &    0.5 & 38.6    & 12 \\ 
NGC4636                 &   323 &   2-7 &    0.0 &   43.5 &      13.5 &    0.2 & 38.3    & 13 \\ 
NGC7507                 & 11344 & 0.3-7 &    0.1 &    7.3 &      18.7 &    2.7 & 39.0    & 14 \\ 
NGC1332                 &  4372 &   2-7 &    0.1 &   31.7 &      28.4 &    0.8 & 39.2    & 15 \\ 
NGC4036                 &  6783 &   2-7 &    0.0 &   13.7 &      83.2 &    6.7 & 40.3    & 11 \\ 
NGC3640                 &  7142 & 0.3-7 &    0.0 &    9.0 &       7.3 & $<$0.7 & $<$38.3 & 11 \\ 
NGC1052                 &  5910 & 0.3-7 &    0.0 &   55.3 &    4952.3 &   94.2 & 40.6    & 3  \\ 
NGC5576                 & 11781 & 0.3-7 &    0.0 &   29.7 &      41.9 &    1.5 & 38.8    & 9  \\ 
NGC4291                 & 11778 &   2-7 &    0.0 &   29.6 &      16.8 &    0.6 & 39.4    & 9  \\ 
NGC5838                 &  6788 &   2-7 &    0.0 &   13.6 &      15.9 &    1.3 & 39.5    & 11 \\ 
NGC5638                 & 11313 & 0.3-7 &    0.0 &   10.0 &       3.0 &   $<$0.4 & $<$38.3 & 14 \\ 
NGC5831                 & 11314 & 0.3-7 &    3.5 &    9.6 &      36.7 &    3.8 & 39.4    & 14 \\ 
NGC3384                 & 11782 &   2-7 &    0.0 &   28.6 &      20.9 &    0.8 & 38.8    & 9  \\ 
NGC3115                 & 12095 &   2-7 &    0.0 &   76.0 &      80.1 &    1.2 & 38.6    & 16 \\ 
NGC3193                 & 11360 & 0.3-7 &    0.0 &    7.2 &      18.6 &    2.7 & 39.0    & 14 \\ 
NGC4203                 & 10535 &   2-7 &    0.0 &   41.4 &    2022.2 &   54.2 & 40.9    & 17 \\ 
NGC1439                 & 11346 & 0.3-7 &    0.0 &    7.5 &      19.0 &    2.6 & 39.1    & 14 \\ 
NGC5582                 & 11361 & 0.3-7 &    0.0 &    9.5 &       9.3 &    1.0 & 38.7    & 14 \\ 
NGC1340                 & 11345 & 0.3-7 &    0.0 &    2.9 &       2.1 & $<$1.1 & $<$38.2 & 14 \\ 
NGC4278                 &  7081 &   2-7 &    0.0 &  110.1 &     966.4 &    9.7 & 39.8    & 18 \\ 
NGC4742                 & 11779 &   2-7 &    0.0 &   32.8 &      37.0 &    1.2 & 39.3    & 9  \\ 
NGC2778                 & 11777 & 0.3-7 &    0.0 &   29.3 &      28.9 &    1.0 & 38.9    & 9  \\ 
NGC4648                 & 11362 & 0.3-7 &    0.0 &   10.8 &      15.5 &    1.5 & 39.0    & 14 \\ 
NGC3377                 &  2934 &   2-7 &    0.2 &   37.6 &      24.0 &    0.7 & 38.6    & 19 \\ 
NGC1426                 & 11347 & 0.3-7 &    0.0 &    5.4 &       2.8 & $<$0.6 & $<$38.3 & 14 \\ 
NGC1172                 & 11348 & 0.3-7 &    0.0 &    6.5 &       4.0 &    0.6 & 38.4    & 14 \\ 
NGC6017                 & 11363 & 0.3-7 &    0.1 &   12.0 &      29.5 &    2.6 & 39.3    & 14 \\ 
NGC5845                 &  4009 &   2-7 &    0.0 &   29.8 &      23.9 &    0.9 & 39.4    & 20 \\ 
NGC3457                 & 11364 & 0.3-7 &    0.1 &    5.2 &       7.8 &    1.5 & 38.7    & 14 \\ 
ESO576-076              & 11316 & 0.3-7 &    0.0 &    8.8 &       1.0 & $<$0.4 & $<$38.4 & 14 \\ 
NGC4283                 &  7081 &   2-7 &    3.5 &  106.7 &      52.7 &    0.5 & 38.9    & 18 \\ 
NGC3928                 & 11365 & 0.3-7 &    0.0 &    4.5 &       5.7 &    1.3 & 38.6    & 14 \\ 
NGC3641                 &  7142 & 0.3-7 &    2.5 &    8.8 &       8.0 &    1.0 & 38.8    &    \\ 
NGC4121                 &  2071 & 0.3-7 &    3.8 &   51.6 &      17.0 &    0.3 & 38.0    &    \\ 
UGC07767                & 11367 & 0.3-7 &    0.0 &    8.5 &       5.6 &    0.7 & 38.5    & 14 \\ 
NGC3265                 & 11368 & 0.3-7 &    0.0 &    8.3 &      17.2 &    2.2 & 39.0    & 14 \\ 
IC1729                  & 11349 & 0.3-7 &    0.0 &    5.5 &      17.0 &    3.2 & 39.0    & 14 \\ 
NGC2970                 & 11369 & 0.3-7 &    0.0 &   10.5 &       8.0 &    0.8 & 38.7    & 14 \\ 
UGC05955                & 11370 & 0.3-7 &    0.0 &    7.4 &       3.4 & $<$0.5 & $<$38.3 & 14 \\ 
NGC3073                 &  7851 & 0.3-7 &    0.0 &    4.7 &       1.0 & $<$0.8 & $<$38.4 &    \\ 
NGC3522                 & 11371 & 0.3-7 &    0.0 &    5.9 &       5.8 &    1.0 & 38.5    & 14 \\ 
NGC1097A                &  1611 & 0.3-7 &    3.4 &    5.2 &       1.0 & $<$0.7 & $<$38.0 & 21 \\ 
NGC4627                 &   797 & 0.3-7 &    2.5 &   56.4 &       5.0 & $<$0.1 & $<$36.9 &    \\ 
NGC1370                 & 11350 & 0.3-7 &    0.0 &    2.7 &       8.0 &    3.1 & 38.7    & 14 \\ 
NGC0855                 &  9550 &   2-7 &    0.0 &   58.6 &      46.9 &    0.9 & 38.6    &    \\ 
NGC7077                 &  7854 & 0.3-7 &    0.0 &    5.1 &       0.0 & $<$0.7 & $<$38.3 &    \\ 
PGC056821               & 11373 & 0.3-7 &    0.0 &   11.5 &       6.7 &    0.6 & 38.6    & 14 \\ 
IC0225                  & 11351 & 0.3-7 &    0.0 &    7.5 &       2.4 & $<$0.5 & $<$38.3 & 14 \\ 
NGC1331                 &  4372 & 0.3-7 &    2.9 &   33.2 &      16.5 &    0.5 & 37.9    & 15 \\ 
ESO540-014              & 11352 &   2-7 &    0.0 &    7.9 &      33.1 &    4.7 & 40.1    & 14 \\ 
PGC042748               & 11318 & 0.3-7 &    0.0 &    3.5 &       0.0 & $<$1.0 & $<$38.3 & 14 \\ 
NGC4308                 &  7853 & 0.3-7 &    0.0 &    4.7 &       2.9 & $<$0.8 & $<$37.9 &    \\ 
NGC5099                 & 11319 & 0.3-7 &    0.0 &    5.8 &       0.0 & $<$0.6 & $<$38.3 & 14 \\ 
PGC042173               & 11320 & 0.3-7 &    0.0 &    7.9 &       0.0 & $<$0.4 & $<$38.3 & 14 \\ 
PGC028305               & 11376 & 0.3-7 &    0.0 &    8.2 &       0.0 & $<$0.4 & $<$38.3 & 14 \\ 
PGC1242097              & 11375 & 0.3-7 &    0.0 &   11.6 &       0.0 & $<$0.3 & $<$38.3 & 14 \\ 
PGC3119319              &   788 & 0.3-7 &    3.1 &   22.2 &       5.0 & $<$0.3 & $<$38.1 &    \\ 
PGC132768               & 11354 & 0.3-7 &    0.0 &    6.6 &       1.0 & $<$0.5 & $<$38.3 & 14 \\ 
PGC1210284              & 11377 & 0.3-7 &    0.0 &   11.2 &       1.0 & $<$0.3 & $<$38.3 & 14 \\ 
PGC030133               & 11378 & 0.3-7 &    0.0 &    4.7 &       0.0 & $<$0.7 & $<$38.3 & 14 \\ 
PGC085239               & 11341 & 0.3-7 &    0.0 &    6.9 &       0.0 & $<$0.5 & $<$38.3 & 14 \\ 
PGC042737               & 11322 & 0.3-7 &    0.0 &   10.7 &       0.0 & $<$0.3 & $<$38.3 & 14 \\ 
PGC3097911              & 11355 & 0.3-7 &    0.0 &    5.7 &       0.0 & $<$0.6 & $<$38.3 & 14 \\ 
PGC064718               & 11342 & 0.3-7 &    0.0 &    2.0 &       1.0 & $<$1.6 & $<$38.4 & 14 \\ 
PGC1209872              & 11379 & 0.3-7 &    0.0 &   11.1 &       1.4 & $<$0.3 & $<$38.3 & 14 \\ 
PGC740586               & 11344 & 0.3-7 &    4.5 &    7.0 &       0.8 & $<$0.5 & $<$38.2 & 14 \\ 
PGC1202458              & 11380 & 0.3-7 &    0.0 &   10.9 &       0.0 & $<$0.3 & $<$38.3 & 14 \\ 
PGC1216386              & 11381 & 0.3-7 &    0.0 &   11.4 &       0.0 & $<$0.3 & $<$38.3 & 14 \\ 
PGC1230503              & 11382 & 0.3-7 &    0.0 &   11.8 &       0.0 & $<$0.3 & $<$38.3 & 14 \\ 
SDSSJ150907.83+004329.7 & 11383 & 0.3-7 &    0.0 &    9.4 &       0.0 & $<$0.4 & $<$38.4 & 14 \\ 
6dFJ2049400-324154      & 11357 & 0.3-7 &    0.0 &    9.0 &       0.0 & $<$0.4 & $<$38.3 & 14 \\ 
SDSSJ150033.02+021349.1 & 11324 & 0.3-7 &    0.0 &    6.3 &       0.0 & $<$0.5 & $<$38.3 & 14 \\ 
SDSSJ145828.64+013234.6 & 11325 & 0.3-7 &    0.0 &    8.4 &       0.0 & $<$0.4 & $<$38.3 & 14 \\ 
SDSSJ150100.85+010049.8 & 11327 & 0.3-7 &    0.0 &   11.4 &       0.0 & $<$0.3 & $<$38.3 & 14 \\ 
SDSSJ150812.35+012959.7 & 11384 & 0.3-7 &    0.0 &   10.0 &       1.0 & $<$0.4 & $<$38.3 & 14 \\ 
PGC1179083              & 11330 & 0.3-7 &    0.1 &    9.9 &       0.0 & $<$0.4 & $<$38.3 & 14 \\ 
PGC1206166              & 11385 & 0.3-7 &    0.0 &   10.5 &       7.0 &    0.7 & 38.6    & 14 \\ 
SDSSJ150233.03+015608.3 & 11329 & 0.3-7 &    0.0 &   10.0 &       0.0 & $<$0.3 & $<$38.3 & 14 \\ 
PGC042724               & 11331 & 0.3-7 &    0.0 &    1.8 &       0.0 & $<$1.8 & $<$38.2 & 14 \\ 
PGC135659               & 11358 & 0.3-7 &    0.0 &    2.7 &       0.0 & $<$1.2 & $<$38.3 & 14 \\ 
PGC135829               & 11333 & 0.3-7 &    0.1 &    6.7 &       0.0 & $<$0.5 & $<$38.3 & 14 \\ 
PGC135818               & 11334 & 0.3-7 &    0.0 &    3.5 &       0.0 & $<$0.9 & $<$38.3 & 14 \\ 
PGC042596               & 11335 & 0.3-7 &    0.0 &    2.5 &       1.0 & $<$1.3 & $<$38.3 & 14 \\ 
PGC1223766              & 11337 & 0.3-7 &    0.0 &    9.2 &       3.2 & $<$0.4 & $<$38.3 & 14 \\ 
PGC043421               & 11338 & 0.3-7 &    0.0 &    3.9 &       0.0 & $<$0.9 & $<$38.3 & 14 \\ 
SDSSJ145944.77+020752.1 & 11336 & 0.3-7 &    0.0 &    8.0 &       0.0 & $<$0.4 & $<$38.3 & 14 \\ 
PGC1192611              & 11339 & 0.3-7 &    0.1 &    8.4 &       0.0 & $<$0.4 & $<$38.3 & 14 \\ 
PGC042454               & 11340 & 0.3-7 &    0.0 &    2.6 &       0.0 & $<$1.3 & $<$38.2 & 14 \\ 
PGC1217593              & 11386 & 0.3-7 &    0.0 &    4.6 &       0.0 & $<$0.7 & $<$38.3 & 14 \\[-17pt] 

\enddata

\tablecomments{Names and row ordering match Table~1. Aperture
photometry is carried out as described in $\S$2 and $L_{\rm X}$ is
calculated as described in $\S$3.1. Snapshot observations are those
for which the reference is 14.\\[-3pt]}

\tablenotetext{a}{References: (1) Zhang \& Xu~(2004); (2) Balmaverde
et al.~(2008); (3) Boroson et al.~(2011); (4) Trinchieri \&
Goudfrooij~(2002); (5) Kim \& Fabbiano~(2010); (6) Sivakoff et
al.~(2005); (7) Irwin et al.~(2003); (8) Fabbiano et al.~(2003); (9)
G{\"u}ltekin et al.~(2012); (10) Humphrey \& Buote~(2010); (11)
Liu~(2011); (12) Fabbiano et al.~(2006); (13) Loewenstein et
al.~(2001); (14) This work; (15) Humphrey \& Buote~(2004); (16) Wong
et al.~(2011); (17) Dunn et al.~(2010); (18) Brassington et
al.~(2009); (19) Irwin et al.~(2004); (20) Soria et al.~(2006a); (21)
Nemmen et al.~(2006).}

\end{deluxetable}

\clearpage

\begin{deluxetable}{p{45pt}rrrrrcp{45pt}rrrrr}
\tablecaption{Sample Properties}
\tablewidth{17.5cm}

\tablehead{\colhead{Sample} & \colhead{$n$} & \colhead{Mean} &
\colhead{25th} & \colhead{50th} & \colhead{75th} & & \colhead{Sample}
& \colhead{$n$} & \colhead{Mean} & \colhead{25th} & \colhead{50th} &
\colhead{75th}\\[-3pt]}

\startdata
\multicolumn{6}{c}{$\log{(M_{\rm star}/M_{\odot})}$}                       &~&  \multicolumn{6}{c}{$M_{\rm B}$ (mag)}   \\               		      
Field                 &  103 &   9.66$\pm$0.12  &  8.56  &  9.75  & 10.84  &~&	 Field                 &  103 & $-$17.55$\pm$0.25  & $-$19.82 & $-$17.64 &$-$15.39  \\ 
~~Snapshot            &   61 &   8.98$\pm$0.13  &  8.02  &  8.71  &  9.96  &~&	 ~~Snapshot            &   61 & $-$16.18$\pm$0.27  & $-$17.84 & $-$15.59 &$-$14.21  \\ 
~~Elliptical          &   93 &   9.62$\pm$0.13  &  8.44  &  9.71  & 10.74  &~&	 ~~Elliptical          &   93 & $-$17.48$\pm$0.26  & $-$19.73 & $-$17.41 &$-$15.22  \\ 
~~In group            &   74 &   9.86$\pm$0.15  &  8.71  & 10.07  & 11.05  &~&	 ~~In group            &   74 & $-$17.94$\pm$0.30  & $-$20.15 & $-$18.21 &$-$15.36  \\[+3pt] 
									   									     
\multicolumn{6}{c}{$\log{(M_{\rm BH}/M_{\odot})}$}                          &~&  \multicolumn{6}{c}{$\log{L_{\rm X}}$ (detections)} \\   		     
Field                 &  103 &   6.90$\pm$0.12  &  5.75  &  6.79  &  7.96  &~&	 Field                 &  50 &  39.18$\pm$0.09 &  38.65 &  39.00  & 39.40 \\	 
~~Snapshot            &   61 &   6.30$\pm$0.12  &  5.50  &  6.08  &  7.08  &~&	 ~~~Snapshot           &  19 &  38.87$\pm$0.09 &  38.58 &  38.69  & 39.03 \\	 
~~Elliptical          &   93 &   6.86$\pm$0.12  &  5.70  &  6.76  &  7.95  &~&	 ~~~Elliptical         &  44 &  39.12$\pm$0.09 &  38.64 &  38.96  & 39.35 \\	 
~~In group            &   74 &   7.08$\pm$0.15  &  5.82  &  7.27  &  8.28  &~&	 ~~~In group           &  38 &  39.22$\pm$0.10 &  38.72 &  39.04  & 39.46 \\[+3pt] 	 

\multicolumn{6}{c}{$\log{(M_{\rm BH}/M_{\rm star})}$}                              &~&  \multicolumn{6}{c}{$\log{(L_{\rm X}/10^{37})}$ (all)\tablenotemark{a}} \\		      
Field                 &  103 &  $-$2.76$\pm$0.04  & $-$2.90  & $-$2.75  & $-$2.53  &~&  Field                  &  97 &  1.71$\pm$0.07 & 0.63 & 1.37 & 2.02  \\	   
~~Snapshot            &   61 &  $-$2.68$\pm$0.05  & $-$2.82  & $-$2.68  & $-$2.48  &~&	 ~~~Snapshot           &  61 &  1.42$\pm$0.05 & 0.44 & 0.88 & 1.55  \\    
~~Elliptical          &   93 &  $-$2.76$\pm$0.04  & $-$2.89  & $-$2.75  & $-$2.55  &~&	 ~~~Elliptical         &  87 &  1.67$\pm$0.07 & 0.62 & 1.32 & 1.97  \\	   
~~In group            &   74 &  $-$2.77$\pm$0.05  & $-$2.89  & $-$2.74  & $-$2.47  &~&	 ~~~In group           &  68 &  1.78$\pm$0.08 & 0.70 & 1.62 & 2.11  \\[-17pt]

\enddata

\tablecomments{Percentiles are given as the value for the nearest
object in the sorted list for detected quantities, and calculated
within ASURV (Kaplan-Meier) for $\log{(L_{\rm X}/10^{37})}$ with upper
limits included; note the 25th percentile values here are dominated by
censored points and should be taken as indications only. The
\hbox{X-ray} luminosities are restricted to $\log{L_{\rm X}}>38.2$ to
avoid any potential bias from archival observations.\\[-3pt]}

\tablenotetext{a}{Values are biased because the ASURV changed the
first upper limit to a detection for the Kaplan-Meier computation.}

\end{deluxetable}

\begin{deluxetable}{p{50pt}rrrrcrrrr}
\tablecaption{Correlation between nuclear X-ray luminosity and stellar mass}
\tablewidth{14.5cm}

\tablehead{ & \multicolumn{4}{c}{All data} & &
\multicolumn{4}{c}{Detections} \\[+3pt] \colhead{Sample} & \colhead{$n$} &
\colhead{$A$} & \colhead{$B$} & \colhead{$\sigma$} & & \colhead{$n$} &
\colhead{$A$} & \colhead{$B$} & \colhead{$\sigma$}\\[-3pt]}

\startdata
Field                 &   97 & $-0.04^{+0.11}_{-0.12}$ & $0.71^{+0.10}_{-0.09}$ & $0.73^{+0.10}_{-0.08}$ &~~&  50  &  $ 0.52^{+0.13}_{-0.12}$ & $0.35^{+0.11}_{-0.12}$ & $0.59^{+0.07}_{-0.06}$  \\[+2pt] 
~~Snapshot            &   61 & $-0.13^{+0.14}_{-0.18}$ & $0.66^{+0.19}_{-0.15}$ & $0.74^{+0.17}_{-0.13}$ &~~&  19  &  $ 0.47^{+0.11}_{-0.10}$ & $0.00^{+0.15}_{-0.14}$ & $0.43^{+0.10}_{-0.08}$  \\[+2pt] 
~~Elliptical          &   87 & $-0.03^{+0.10}_{-0.11}$ & $0.65^{+0.09}_{-0.08}$ & $0.71^{+0.09}_{-0.08}$ &~~&  44  &  $ 0.51^{+0.11}_{-0.11}$ & $0.30^{+0.11}_{-0.11}$ & $0.54^{+0.07}_{-0.06}$  \\[+2pt] 
~~In group            &   68 & $-0.07^{+0.13}_{-0.13}$ & $0.72^{+0.11}_{-0.11}$ & $0.67^{+0.10}_{-0.08}$ &~~&  38  &  $ 0.43^{+0.15}_{-0.15}$ & $0.43^{+0.13}_{-0.13}$ & $0.54^{+0.08}_{-0.06}$  \\[-17pt]
\enddata

\tablecomments{Fits are to the function $(\log{L_{\rm
x}}-38.4)=A+B{\times}(\log{M_{\rm star}}-9.8)$, where $\sigma$ is the
intrinsic scatter. Fitting is carried out using the IDL code of
Kelly~(2007); see $\S$4.2 for details.}

\end{deluxetable}


\begin{thebibliography}

\small


\bibitem[Acreman et al.(2003)]{2003MNRAS.341.1333A} Acreman, D.~M.,
  Stevens, I.~R., Ponman, T.~J., \& Sakelliou, I.\ 2003, \mnras, 341,
  1333

\bibitem[Baganoff et al.(2001)]{2001Natur.413...45B} Baganoff, F.~K.,
  et al.\ 2001, \nat, 413, 45

\bibitem[Balmaverde et al.(2008)]{2008A&A...486..119B} Balmaverde, B., 
  Baldi, R.~D., \& Capetti, A.\ 2008, \aap, 486, 119

\bibitem[Bauer et al.(2004)]{2004AJ....128.2048B} Bauer, F.~E., Alexander, 
D.~M., Brandt, W.~N., et al.\ 2004, \aj, 128, 2048 

\bibitem[Begelman(2010)]{2010MNRAS.402..673B} Begelman, M.~C.\ 2010,
  \mnras, 402, 673

\bibitem[Bekki \& Graham(2010)]{2010ApJ...714L.313B} Bekki, K., \&
  Graham, A.~W.\ 2010, \apjl, 714, L313

\bibitem[Bell \& de Jong(2001)]{2001ApJ...550..212B} Bell, E.~F., \&
  de Jong, R.~S.\ 2001, \apj, 550, 212

\bibitem[Bell et al.(2003)]{2003ApJS..149..289B} Bell, E.~F.,
  McIntosh, D.~H., Katz, N., \& Weinberg, M.~D.\ 2003, \apjs, 149, 289

\bibitem[Bloom et al.(2011)]{2011Sci...333..203B} Bloom, J.~S., et
  al.\ 2011, Science, 333, 203

\bibitem[Boroson et al.(2011)]{2011ApJ...729...12B} Boroson, B., Kim,
  D.-W., \& Fabbiano, G.\ 2011, \apj, 729, 12

\bibitem[Brassington et al.(2009)]{2009ApJS..181..605B} Brassington, N.~J., 
  Fabbiano, G., Kim, D.-W., et al.\ 2009, \apjs, 181, 605 

\bibitem[Brockamp et al.(2011)]{2011MNRAS.418.1308B} Brockamp, M., 
Baumgardt, H., \& Kroupa, P.\ 2011, \mnras, 418, 1308 

\bibitem[Brown \& Bregman(2000)]{2000ApJ...539..592B} Brown, B.~A., \&
  Bregman, J.~N.\ 2000, \apj, 539, 592

\bibitem[Brown et al.(2006)]{2006ApJ...638...88B} Brown, M.~J.~I., Brand, 
K., Dey, A., et al.\ 2006, \apj, 638, 88 

\bibitem[Buckley \& James(1979)]{BJ79} Buckley, J., \& James, I., 1979, 
   Biometrics, 66, 429

\bibitem[Burrows et al.(2011)]{2011Natur.476..421B} Burrows, D.~N., Kennea, 
  J.~A., Ghisellini, G., et al.\ 2011, \nat, 476, 421

\bibitem[Cannizzo et al.(1990)]{1990ApJ...351...38C} Cannizzo, J.~K., Lee, 
  H.~M., \& Goodman, J.\ 1990, \apj, 351, 38 

\bibitem[Cannizzo et al.(2011)]{2011ApJ...742...32C} Cannizzo, J.~K., 
Troja, E., \& Lodato, G.\ 2011, \apj, 742, 32 

\bibitem[Choi et al.(2009)]{2009ApJ...699.1679C} Choi, Y.-Y., Woo, J.-H.,
  \& Park, C.\ 2009, \apj, 699, 1679

\bibitem[Ciotti et al.(2009)]{2009ApJ...699...89C} Ciotti, L.,
  Ostriker, J.~P., \& Proga, D.\ 2009, \apj, 699, 89

\bibitem[Ciotti et al.(2010)]{2010ApJ...717..708C} Ciotti, L.,
  Ostriker, J.~P., \& Proga, D.\ 2010, \apj, 717, 708

\bibitem[C{\^o}t{\'e} et al.(2004)]{2004ApJS..153..223C} C{\^o}t{\'e},
P., et al.\ 2004, \apjs, 153, 223

\bibitem[Croom et al.(2009)]{2009MNRAS.399.1755C} Croom, S.~M., Richards, 
G.~T., Shanks, T., et al.\ 2009, \mnras, 399, 1755 

\bibitem[Croton et al.(2006)]{2006MNRAS.365...11C} Croton, D.~J., et
  al.\ 2006, \mnras, 365, 11

\bibitem[De Lucia et al.(2006)]{2006MNRAS.366..499D} De Lucia, G.,
  Springel, V., White, S.~D.~M., Croton, D., \& Kauffmann, G.\ 2006,
  \mnras, 366, 499

\bibitem[Di Matteo et al.(2000)]{2000MNRAS.311..507D} Di Matteo, T.,
  Quataert, E., Allen, S.~W., Narayan, R., \& Fabian, A.~C.\ 2000,
  \mnras, 311, 507

\bibitem[Dunn et al.(2010)]{2010MNRAS.404..180D} Dunn, R.~J.~H., Allen, 
  S.~W., Taylor, G.~B., et al.\ 2010, \mnras, 404, 180 

\bibitem[Eckart et al.(2005)]{2005ApJS..156...35E} Eckart, M.~E.,
  Laird, E.~S., Stern, D., Mao, P.~H., Helfand, D.~J., \& Harrison,
  F.~A.\ 2005, \apjs, 156, 35

\bibitem[Eliche-Moral et al.(2010)]{2010A&A...519A..55E} Eliche-Moral,
  M.~C., et al.\ 2010, \aap, 519, A55

\bibitem[Fabbiano et al.(2003)]{2003ApJ...588..175F} Fabbiano, G., Elvis, 
  M., Markoff, S., et al.\ 2003, \apj, 588, 175

\bibitem[Fabbiano et al.(2006)]{2006ApJ...650..879F} Fabbiano, G., Kim, 
  D.-W., Fragos, T., et al.\ 2006, \apj, 650, 879 

\bibitem[Ferrarese \& Ford(2005)]{2005SSRv..116..523F} Ferrarese, L., 
   \& Ford, H.\ 2005, \ssr, 116, 523 

\bibitem[Ferrarese et al.(2006)]{2006ApJ...644L..21F} Ferrarese, L., 
   C{\^o}t{\'e}, P., Dalla Bont{\`a}, E., et al.\ 2006a, \apjl, 644, L21 

\bibitem[Ferrarese et al.(2006)]{2006ApJS..164..334F} Ferrarese, L., 
  C{\^o}t{\'e}, P., Jord{\'a}n, A., et al.\ 2006b, \apjs, 164, 334 

\bibitem[Filho et al.(2004)]{2004A&A...418..429F} Filho, M.~E., 
  Fraternali, F., Markoff, S., et al.\ 2004, \aap, 418, 429 

\bibitem[Fukugita et al.(1996)]{1996AJ....111.1748F} Fukugita, M.,
  Ichikawa, T., Gunn, J.~E., Doi, M., Shimasaku, K., \& Schneider,
  D.~P.\ 1996, \aj, 111, 1748

\bibitem[Gallo et al.(2008)]{2008ApJ...680..154G} Gallo, E., Treu, T.,
  Jacob, J., Woo, J.-H., Marshall, P.~J., \& Antonucci, R.\ 2008,
  \apj, 680, 154

\bibitem[Gallo et al.(2010)]{2010ApJ...714...25G} Gallo, E., Treu, T.,
  Marshall, P.~J., Woo, J.-H., Leipski, C., \& Antonucci, R.\ 2010,
  \apj, 714, 25

\bibitem[Gavazzi et al.(2010)]{2010A&A...517A..73G} Gavazzi, G.,
  Fumagalli, M., Cucciati, O., \& Boselli, A.\ 2010, \aap, 517, A73

\bibitem[Gehrels(1986)]{1986ApJ...303..336G} Gehrels, N.\ 1986, 
\apj, 303, 336 

\bibitem[Ghosh et al.(2009)]{2009AJ....137.3263G} Ghosh, K.~K., Saripalli, 
  L., Gandhi, P., Foellmi, C., Guti{\'e}rrez, C.~M., \&
  L{\'o}pez-Corredoira, M.\ 2009, \aj, 137, 3263

\bibitem[Gilfanov(2004)]{2004MNRAS.349..146G} Gilfanov, M.\ 2004, 
\mnras, 349, 146 

\bibitem[Gnedin(2003)]{2003ApJ...582..141G} Gnedin, O.~Y.\ 2003, \apj, 582, 
  141

\bibitem[Gonz{\'a}lez-Mart{\'{\i}}n et al.(2009)]{2009ApJ...704.1570G} 
Gonz{\'a}lez-Mart{\'{\i}}n, O., Masegosa, J., M{\'a}rquez, I., 
\& Guainazzi, M.\ 2009, \apj, 704, 1570 

\bibitem[Goulding et al.(2010)]{2010MNRAS.406..597G} Goulding, A.~D.,
  Alexander, D.~M., Lehmer, B.~D., \& Mullaney, J.~R.\ 2010, \mnras,
  406, 597

\bibitem[Graham(2008)]{2008ApJ...680..143G} Graham, A.~W.\ 2008, 
\apj, 680, 143 

\bibitem[Graham \& Spitler(2009)]{2009MNRAS.397.2148G} Graham, A.~W.,
  \& Spitler, L.~R.\ 2009, \mnras, 397, 2148

\bibitem[Graham et al.(2011)]{2011MNRAS.412.2211G} Graham, A.~W.,
  Onken, C.~A., Athanassoula, E., \& Combes, F.\ 2011, \mnras, 412,
  2211

\bibitem[Greene et al.(2010)]{2010ApJ...721...26G} Greene, J.~E., Peng, 
C.~Y., Kim, M., et al.\ 2010, \apj, 721, 26 

\bibitem[G{\"u}ltekin et al.(2009)]{2009ApJ...698..198G} G{\"u}ltekin,
  K., et al.\ 2009, \apj, 698, 198

\bibitem[G{\"u}ltekin et al.(2012)]{xcombh} G{\"u}ltekin, K., Cackett, 
E.~M., Miller, J.~M., et al.\ 2012, \apj, submitted

\bibitem[Hasinger et al.(2005)]{2005A&A...441..417H} Hasinger, G., 
  Miyaji, T., \& Schmidt, M.\ 2005, \aap, 441, 417

\bibitem[Healey et al.(2007)]{2007ApJS..171...61H} Healey, S.~E.,
  Romani, R.~W., Taylor, G.~B., Sadler, E.~M., Ricci, R., Murphy, T.,
  Ulvestad, J.~S., \& Winn, J.~N.\ 2007, \apjs, 171, 61

\bibitem[Heckman et al.(2004)]{2004ApJ...613..109H} Heckman, T.~M., 
  Kauffmann, G., Brinchmann, J., Charlot, S., Tremonti, C., \& White,
  S.~D.~M.\ 2004, \apj, 613, 109

\bibitem[Ho et al.(1997)]{1997ApJ...487..591H} Ho, L.~C., Filippenko,
  A.~V., \& Sargent, W.~L.~W.\ 1997, \apj, 487, 591

\bibitem[Ho(1999)]{1999ApJ...516..672H} Ho, L.~C.\ 1999, \apj, 516, 672 

\bibitem[Ho(2008)]{2008ARA&A..46..475H} Ho, L.~C.\ 2008, \araa, 46, 475 

\bibitem[Hong et al.(2005)]{2005ApJ...635..907H} Hong, J., van den
  Berg, M., Schlegel, E.~M., Grindlay, J.~E., Koenig, X., Laycock, S.,
  \& Zhao, P.\ 2005, \apj, 635, 907

\bibitem[Hopkins et al.(2005)]{2005ApJ...625L..71H} Hopkins, P.~F., 
Hernquist, L., Martini, P., et al.\ 2005, \apjl, 625, L71 

\bibitem[Hopkins et al.(2007)]{2007ApJ...654..731H} Hopkins, P.~F., 
  Richards, G.~T., \& Hernquist, L.\ 2007, \apj, 654, 731

\bibitem[Hopkins \& Quataert(2011)]{2011MNRAS.411L..61H} Hopkins,
  P.~F., \& Quataert, E.\ 2011, \mnras, 411, L61

\bibitem[Humphrey \& Buote(2004)]{2004ApJ...612..848H} Humphrey, P.~J., 
  \& Buote, D.~A.\ 2004, \apj, 612, 848 

\bibitem[Humphrey \& Buote(2010)]{2010MNRAS.403.2143H} Humphrey, P.~J., 
  \& Buote, D.~A.\ 2010, \mnras, 403, 2143 

\bibitem[Irwin et al.(2003)]{2003ApJ...587..356I} Irwin, J.~A., Athey, 
  A.~E., \& Bregman, J.~N.\ 2003, \apj, 587, 356

\bibitem[Irwin et al.(2004)]{2004ApJ...601L.143I} Irwin, J.~A., Bregman, 
  J.~N., \& Athey, A.~E.\ 2004, \apjl, 601, L143 

\bibitem[Jiang et al.(2010)]{2010Natur.464..380J} Jiang, L., et al.\
  2010, \nat, 464, 380

\bibitem[Kang et al.(2007)]{2007MNRAS.381..389K} Kang, X., van den
  Bosch, F.~C., \& Pasquali, A.\ 2007, \mnras, 381, 389

\bibitem[Kaplan \& Meier(1968)]{KM68} Kaplan, E. L., \& Meier, P., 1968, 
  JASA, 53, 457

\bibitem[Kelly(2007)]{2007ApJ...665.1489K} Kelly, B.~C.\ 2007, \apj,
  665, 1489

\bibitem[Kelly et al.(2010)]{2010ApJ...719.1315K} Kelly, B.~C.,
  Vestergaard, M., Fan, X., Hopkins, P., Hernquist, L., \&
  Siemiginowska, A.\ 2010, \apj, 719, 1315

\bibitem[Kim \& Fabbiano(2010)]{2010ApJ...721.1523K} Kim, D.-W., 
  \& Fabbiano, G.\ 2010, \apj, 721, 1523

\bibitem[King et al.(2001)]{2001ApJ...552L.109K} King, A.~R., Davies,
  M.~B., Ward, M.~J., Fabbiano, G., \& Elvis, M.\ 2001, \apjl, 552,
  L109

\bibitem[Kormendy et al.(2009)]{2009ApJS..182..216K} Kormendy, J.,
  Fisher, D.~B., Cornell, M.~E., \& Bender, R.\ 2009, \apjs, 182, 216

\bibitem[Kraft et al.(1991)]{1991ApJ...374..344K} Kraft, R.~P., Burrows, 
D.~N., \& Nousek, J.~A.\ 1991, \apj, 374, 344 

\bibitem[Kreckel et al.(2011)]{2011AJ....141....4K} Kreckel, K., et
  al.\ 2011, \aj, 141, 4

\bibitem[Lamastra et al.(2010)]{2010MNRAS.405...29L} Lamastra, A.,
  Menci, N., Maiolino, R., Fiore, F., \& Merloni, A.\ 2010, \mnras,
  405, 29

\bibitem[Latta(1981)]{L81} Latta, R.~P., 1981, JASA, 76, 713

\bibitem[Lauer et al.(2007)]{2007ApJ...670..249L} Lauer, T.~R.,
  Tremaine, S., Richstone, D., \& Faber, S.~M.\ 2007, \apj, 670, 249

\bibitem[Lavalley et al.(1992)]{1992ASPC...25..245L} Lavalley, M., Isobe, 
  T., \& Feigelson, E.\ 1992, Astronomical Data Analysis Software and
  Systems I, 25, 245

\bibitem[Leipski et al.(2012)]{2012ApJ...744..152L} Leipski, C., Gallo, E., 
Treu, T., et al.\ 2012, \apj, 744, 152 

\bibitem[Levan et al.(2011)]{2011Sci...333..199L} Levan, A.~J., Tanvir, 
  N.~R., Cenko, S.~B., et al.\ 2011, Science, 333, 199 

\bibitem[Li et al.(2002)]{2002ApJ...576..753L} Li, L.-X., Narayan, R., 
  \& Menou, K.\ 2002, \apj, 576, 753 

\bibitem[Lin et al.(2011)]{2011ApJ...738...52L} Lin, D., Carrasco,
  E.~R., Grupe, D., Webb, N.~A., Barret, D., \& Farrell, S.~A.\ 2011,
  \apj, 738, 52

\bibitem[Liu(2011)]{2011ApJS..192...10L} Liu, J.\ 2011, \apjs, 192, 10 

\bibitem[Lodato et al.(2009)]{2009MNRAS.392..332L} Lodato, G., King, A.~R., 
  \& Pringle, J.~E.\ 2009, \mnras, 392, 332 

\bibitem[Loewenstein et al.(2001)]{2001ApJ...555L..21L} Loewenstein, M., 
  Mushotzky, R.~F., Angelini, L., Arnaud, K.~A., \& Quataert, E.\
  2001, \apjl, 555, L21

\bibitem[Luo et al.(2008)]{2008ApJ...674..122L} Luo, B., Brandt,
  W.~N., Steffen, A.~T., \& Bauer, F.~E.\ 2008, \apj, 674, 122

\bibitem[Makarov \& Karachentsev(2011)]{2011MNRAS.412.2498M} Makarov, D., 
  \& Karachentsev, I.\ 2011, \mnras, 412, 2498 

\bibitem[Marulli et al.(2008)]{2008MNRAS.385.1846M} Marulli, F.,
  Bonoli, S., Branchini, E., Moscardini, L., \& Springel, V.\ 2008,
  \mnras, 385, 1846

\bibitem[Mayer et al.(2010)]{2010Natur.466.1082M} Mayer, L., Kazantzidis, 
S., Escala, A., \& Callegari, S.\ 2010, \nat, 466, 1082 

\bibitem[Mei et al.(2007)]{2007ApJ...655..144M} Mei, S., Blakeslee, J.~P., 
C{\^o}t{\'e}, P., et al.\ 2007, \apj, 655, 144 

\bibitem[Merloni \& Heinz(2007)]{2007MNRAS.381..589M} Merloni, A., \&
Heinz, S.\ 2007, \mnras, 381, 589

\bibitem[Merritt \& Wang(2005)]{2005ApJ...621L.101M} Merritt, D., \&
  Wang, J.\ 2005, \apjl, 621, L101

\bibitem[Miller et al.(2012)]{2012ApJ...745L..13M} Miller, B., Gallo, E., 
Treu, T., \& Woo, J.-H.\ 2012, \apjl, 745, L13 

\bibitem[Miller \& G{\"u}ltekin(2011)]{2011ApJ...738L..13M}
  Miller, J.~M., G{\"u}ltekin, K.\ 2011, \apjl, 738, L13

\bibitem[Moretti et al.(2003)]{2003ApJ...588..696M} Moretti, A., Campana, 
S., Lazzati, D., \& Tagliaferri, G.\ 2003, \apj, 588, 696 

\bibitem[Nemmen et al.(2006)]{2006ApJ...643..652N} Nemmen, R.~S., 
  Storchi-Bergmann, T., Yuan, F., et al.\ 2006, \apj, 643, 652 

\bibitem[Netzer et al.(2007)]{2007ApJ...671.1256N} Netzer, H., Lira,
  P., Trakhtenbrot, B., Shemmer, O., \& Cury, I.\ 2007, \apj, 671,
  1256

\bibitem[Oosterloo et al.(2010)]{2010MNRAS.409..500O} Oosterloo, T.,
  et al.\ 2010, \mnras, 409, 500

\bibitem[O'Sullivan et al.(2001)]{2001MNRAS.328..461O} O'Sullivan, E.,
  Forbes, D.~A., \& Ponman, T.~J.\ 2001, \mnras, 328, 461

\bibitem[Pellegrini(2005)]{2005ApJ...624..155P} Pellegrini, S.\ 2005,
  \apj, 624, 155

\bibitem[Pellegrini(2010)]{2010ApJ...717..640P} Pellegrini, S.\ 2010,
  \apj, 717, 640

\bibitem[Pellegrini et al.(2012)]{2012ApJ...744...21P} Pellegrini, S., 
  Ciotti, L., \& Ostriker, J.~P.\ 2012, \apj, 744, 21 

\bibitem[Paturel et al.(2003)]{2003A&A...412...45P} Paturel G., Petit
  C., Prugniel P., Theureau G., Rousseau J., Brouty M., Dubois P.,
  Cambr{\'e}sy L., 2003, A\&A, 412, 45

\bibitem[Rees(1988)]{1988Natur.333..523R} Rees, M.~J.\ 1988, \nat,
  333, 523

\bibitem[Schawinski et al.(2010)]{2010ApJ...711..284S} Schawinski, K.,
  et al.\ 2010, \apj, 711, 284

\bibitem[Sch{\"o}del et al.(2009)]{2009A&A...502...91S} Sch{\"o}del,
  R., Merritt, D., \& Eckart, A.\ 2009, \aap, 502, 91

\bibitem[Schulze \& Wisotzki(2010)]{2010A&A...516A..87S} Schulze, A., 
\& Wisotzki, L.\ 2010, \aap, 516, A87 

\bibitem[Seth et al.(2008)]{2008ApJ...678..116S} Seth, A., Ag{\"u}eros, M., 
  Lee, D., \& Basu-Zych, A.\ 2008, \apj, 678, 116

\bibitem[Shankar et al.(2009)]{2009ApJ...690...20S} Shankar, F., Weinberg, 
  D.~H., \& Miralda-Escud{\'e}, J.\ 2009, \apj, 690, 20

\bibitem[Shemmer et al.(2006)]{2006ApJ...644...86S} Shemmer, O., et
  al.\ 2006, \apj, 644, 86

\bibitem[Shin et al.(2012)]{2012ApJ...745...13S} Shin, M.-S., Ostriker, 
J.~P., \& Ciotti, L.\ 2012, \apj, 745, 13

\bibitem[Sivakoff et al.(2005)]{2005ApJ...624L..17S} Sivakoff, G.~R., 
  Sarazin, C.~L., \& Jord{\'a}n, A.\ 2005, \apjl, 624, L17

\bibitem[Sivakoff et al.(2007)]{2007ApJ...660.1246S} Sivakoff, G.~R., 
Jord{\'a}n, A., Sarazin, C.~L., et al.\ 2007, \apj, 660, 1246 

\bibitem[Soltan(1982)]{1982MNRAS.200..115S} Soltan, A.\ 1982, \mnras,
  200, 115

\bibitem[Soria et al.(2006a)]{2006ApJ...640..126S} Soria, R., Fabbiano, G., 
  Graham, A.~W., et al.\ 2006a, \apj, 640, 126 

\bibitem[Soria et al.(2006)]{2006ApJ...640..143S} Soria, R., Graham,
  A.~W., Fabbiano, G., Baldi, A., Elvis, M., Jerjen, H., Pellegrini,
  S., \& Siemiginowska, A.\ 2006b, \apj, 640, 143

\bibitem[Strubbe \& Quataert(2009)]{2009MNRAS.400.2070S} Strubbe, L.~E., 
  \& Quataert, E.\ 2009, \mnras, 400, 2070 

\bibitem[Strubbe \& Quataert(2011)]{2011MNRAS.415..168S} Strubbe,
  L.~E., \& Quataert, E.\ 2011, \mnras, 415, 168

\bibitem[Thomas et al.(2005)]{2005ApJ...621..673T} Thomas, D., Maraston, 
  C., Bender, R., \& Mendes de Oliveira, C.\ 2005, \apj, 621, 673

\bibitem[Treu et al.(2003)]{2003ApJ...591...53T} Treu, T., Ellis,
  R.~S., Kneib, J.-P., Dressler, A., Smail, I., Czoske, O., Oemler,
  A., \& Natarajan, P.\ 2003, \apj, 591, 53

\bibitem[Treu et al.(2007)]{2007ApJ...667..117T} Treu, T., Woo, J.-H.,
  Malkan, M.~A., \& Blandford, R.~D.\ 2007, \apj, 667, 117

\bibitem[Treister et al.(2011)]{2011Natur.474..356T} Treister, E., 
  Schawinski, K., Volonteri, M., Natarajan, P., \& Gawiser, E.\ 2011,
  \nat, 474, 356

\bibitem[Trinchieri \& Goudfrooij(2002)]{2002A&A...386..472T} Trinchieri, 
  G., \& Goudfrooij, P.\ 2002, \aap, 386, 472 

\bibitem[V{\'e}ron-Cetty \& V{\'e}ron(2006)]{2006A&A...455..773V}
  V{\'e}ron-Cetty, M.-P., \& V{\'e}ron, P.\ 2006, \aap, 455, 773

\bibitem[Vestergaard \& Osmer(2009)]{2009ApJ...699..800V} Vestergaard,
  M., \& Osmer, P.~S.\ 2009, \apj, 699, 800

\bibitem[Volonteri \& Rees(2006)]{2006ApJ...650..669V} Volonteri, M.,
  \& Rees, M.~J.\ 2006, \apj, 650, 669

\bibitem[Volonteri \& Natarajan(2009)]{2009MNRAS.400.1911V} Volonteri,
  M., \& Natarajan, P.\ 2009, \mnras, 400, 1911

\bibitem[Volonteri(2010)]{2010Natur.466.1049V} Volonteri, M.\ 2010, 
\nat, 466, 1049

\bibitem[Volonteri et al.(2011)]{2011ApJ...730..145V} Volonteri, M.,
  Dotti, M., Campbell, D., \& Mateo, M.\ 2011, \apj, 730, 145

\bibitem[Willott et al.(2010)]{2010AJ....140..546W} Willott, C.~J., et
  al.\ 2010, \aj, 140, 546

\bibitem[Wong et al.(2011)]{2011ApJ...736L..23W} Wong, K.-W., Irwin, J.~A., 
  Yukita, M., et al.\ 2011, \apjl, 736, L23 

\bibitem[Woo et al.(2008)]{2008ApJ...681..925W} Woo, J.-H., Treu, T., 
  Malkan, M.~A., \& Blandford, R.~D.\ 2008, \apj, 681, 925

\bibitem[Woo et al.(2010)]{2010ApJ...716..269W} Woo, J.-H., et al.\
  2010, \apj, 716, 269

\bibitem[Yu \& Tremaine(2002)]{2002MNRAS.335..965Y} Yu, Q., \&
  Tremaine, S.\ 2002, \mnras, 335, 965

\bibitem[Zhang et al.(2009)]{2009ApJ...699..281Z} Zhang, W.~M., Soria,
  R., Zhang, S.~N., Swartz, D.~A., \& Liu, J.~F.\ 2009, \apj, 699, 281

\bibitem[Zhang \& Xu(2004)]{2004ChJAA...4..221Z} Zhang, Z.-L., \& Xu,
  H.-G.\ 2004, Chinese J.~Astron.~Astrophys., 4, 221

\normalsize

\end{thebibliography}
\end{document}